\documentclass[twocolumn]{aastex63}
\usepackage{tabularx}
\usepackage{chngcntr}
\usepackage{mathtools,amssymb}

\newcommand{\ith}{\ensuremath{^{\rm th}}}
\newcommand{\teff}{\ensuremath{T_{\mbox{\scriptsize eff}}}}
\newcommand{\logg}{\ensuremath{\log g}}
\newcommand{\alphafe}{[$\alpha$/Fe]}

\submitjournal{\textit{The Astronomical Journal}}

\graphicspath{{./}{figures/}}

\newcommand{\eg}{{\it e.g.}}

\newcommand{\kepler}{{Kepler}}
\newcommand{\gaia}{{Gaia}}
\newcommand{\tess}{{TESS}}
\newcommand{\ha}{high-$\alpha$}
\newcommand{\highalpha}{high-$\alpha$}
\newcommand{\lowalpha}{low-$\alpha$~}
\newcommand{\dnu}{$\Delta\nu$}
\newcommand{\dP}{$\Delta$P}
\newcommand{\sigmaint}{$\sigma_{int}$}

\shorttitle{Universal properties of the high- and low-$\alpha$ disk}
\shortauthors{Lu et al.}

\begin{document}

\title{Universal properties of the high- and low$-\alpha$ disk: small intrinsic abundance scatter and migrating stars}

\correspondingauthor{Yuxi (Lucy) Lu}
\email{lucylulu12311@gmail.com}

\newcommand{\amnh}{American Museum of Natural History, Central Park West, Manhattan, NY, USA}
\newcommand{\cca}{Center for Computational Astrophysics, Flatiron Institute, 162 5\ith\ Avenue, Manhattan, NY, USA}
\newcommand{\columbia}{Department of Astronomy, Columbia University, 550 West 120\ith\ Street, New York, NY, USA}
\newcommand{\AIP}{Leibniz Institute for Astrophysics Potsdam, An der Sternwarte 16, 14482 Potsdam, Germany}

\author[0000-0003-4769-3273]{Yuxi(Lucy) Lu}
\affiliation{\columbia}
\affiliation{\amnh}

\author{Melissa Ness}
\affiliation{\columbia}

\author{Tobias Buck}
\affiliation{\AIP}

\author{Joel Zinn}
\altaffiliation{NSF Astronomy and Astrophysics Postdoctoral Fellow}
\affiliation{\amnh}

\begin{abstract}
The detailed age-chemical abundance relations of stars measures time-dependent chemical evolution. 
These trends offer strong empirical constraints on nucleosynthetic  processes, as well as the homogeneity of star-forming gas.
Characterizing chemical abundances of stars across the Milky Way over time has been made possible very recently, thanks to surveys like \gaia, APOGEE and \kepler.
Studies of the low-$\alpha$ disk have shown that individual elements have unique age-abundance trends and the intrinsic dispersion around these relations is small.
In this study, we examine and compare the age distribution of stars across both the high and low-$\alpha$ disk and quantify the intrinsic dispersion of 16 elements around their age-abundance relations at [Fe/H] = 0 using APOGEE DR16. 
We find the high-$\alpha$ disk has shallower age-abundance relations compared to the low-$\alpha$ disk, but similar median intrinsic dispersions of $\approx$ 0.04 dex, suggesting universal element production mechanisms for the high and low-$\alpha$ disks, despite differences in formation history.
We visualize the temporal and spatial distribution of disk stars in small chemical cells, revealing signatures of upside-down and inside-out formation.
Further, the metallicity skew and the [Fe/H]-age relations across radius indicates different initial metallicity gradients and evidence for radial migration. 
Our study is accompanied by an age catalogue for 64,317 stars in APOGEE derived using \texttt{The Cannon} with $\approx$ 1.9 Gyr uncertainty across all ages (APO-CAN stars) as well as a red clump catalogue of 22,031 stars with a contamination rate of 2.7\%.

\end{abstract}

\section{Introduction} \label{sec:intro}

Large spectroscopic surveys such as Apache Point Observatory Galactic Evolution Experiment (APOGEE) \citep{Majewski2017}, Large Sky Area Multi-Object Fibre Spectroscopic Telescope (LAMOST) \citep{LAMOST}, GALactic Archaeology with HERMES (GALAH) \citep{Silva2015, Buder2019} and time-domain surveys such as \kepler\ \citep{Borucki2010} and \tess\ \citep{TESS} are observing hundreds of thousands of stars.
These surveys provide the data that gives us insight as to the formation and evolution of the Galaxy as well as the nucleosynthetic channels of chemical enrichment.

The APOGEE survey \citep{Majewski2017} is an IR survey at R=22,500 that primarily targets the disk, where the majority of the baryonic matter of the Milky Way resides \citep[e.g.][]{Bland2016}.
From the APOGEE R=22,500 spectra, more than 20 precision element abundances [X/Fe] \citep{garcia2016, DR162020}, imprecise spectroscopic ages \citep[\eg][]{Leung2019,Ness2016,Martig2015}, and precision distances \citep[e.g.][]{Leung2019, Hogg2019} can be determined.
Time-domain missions, most notably to date the \kepler\ survey \citep{Borucki2010}, are enabling precision ages to be determined via asteroseismology by examining internal oscillation frequencies of stars.
A population of \kepler\ red giants with both asteroseismic data and APOGEE spectra are collated in the APOKASC catalogue \citep{Pinsonneault2018}. This provides 2,616 precise asteroseismic ages. This catalogue has proven to be a useful benchmark for building larger catalogues of stellar ages using machine learning \citep[e.g.,][]{Ness2016,Ness2019,Mackereth2019}.

Using both 1) small local benchmark samples of stars with high precision abundances and ages from stellar spectra and astroseismology and 2) large samples of stars with high precision abundances and imprecise ages across the Galactic disk allows for testing the chemical enrichment of the disk over wide ranges in time and spatial position. 
With these data, we can examine global age distributions of stars across the Milky Way as well as the temporal and spatial properties of stars with different chemical compositions. 
Globally, this can link the star formation history to the galaxy formation history and reveal evolutionary processes at work like radial migration \citep[e.g.][]{Roskar2008}.

The spectroscopic age distributions built using large surveys have shown the detailed mapping from the old populations in the inner Galaxy, to the young populations in the outer disk \citep[e.g.][]{Ness2016, Martig2016, Bovy2019, Bensby2017}. Further, younger stars are clearly concentrated to the plane of the disk and old stars at larger heights with flaring across radius \citep[e.g.][]{Mackereth2019, Martig2016}. 
The age gradient indicates an inside-out formation for the Galaxy.
Although stars also evolve from their birth sites over time, stellar ages have been used to model the so called radial migration across part of the disk, which has been determined to be strong \citep[][]{Frankel2018, Frankel2019}. To connect the star formation environment and history to formation and evolutionary processes like radial migration, we need to explore age-individual chemical abundance relations at different locations of the disk.

Age-individual chemical abundance trends at fixed metallicity find utility as chemical clocks, via which we can understand:

\begin{itemize}
    \item Nucleosynthesis processes: 
    Different elements are believed to be produced in different processes on different timescales. 
    For example, light elements such as C and N are produced in large part during the phase of asymptotic giant branch (AGB) stars; iron-peak elements (e.g. V, Cr, Mn, Ni, Co) are produced mostly by type Ia supernovae; $\alpha$-elements (e.g. O, Mg, Si, S, and Ca) derive from core-collapse supernovae. Many elements are produced by multiple channels are have both both mass and metallicity dependent yields \citep[more detailed description and references see][]{Kobayashi2020}.
    These complicated nucleosynthesis processes and stellar yields are in detail based on many approximations and estimates. By studying the age-chemical abundance trends, one can learn the chemical yields as informed by the data, and constrain the theoretical models \citep{Rybizki2017}. 
    \item Formation processes in the Milky Way:
    By combining the insights from chemical clocks with spatial and kinematic properties of stars across the Galaxy, we can study how the disk has formed and evolved subsequently through radial migration \citep[e.g.][]{Frankel2018, Frankel2019}. Ultimately we can use this information in combination with simulations to link the current day properties of stars to their birth location and environments. 
\end{itemize}


Chemically, the disk is broadly characterised by the presence of a high and low-$\alpha$ sequence of stars. 
The [$\alpha$/Fe]-[Fe/H] bi-modality was discovered by \cite{Fuhrmann1998}.
The stars in the high-$\alpha$ disk are predominantly old and those in the low alpha disk are predominantly young \citep[e.g.][]{Bensby2014}. 
This bi-modality has been linked to the structural ``thin'' and ``thick'' disk \cite{Gilmore1983}, and certainly, the [$\alpha$-Fe] versus [Fe/H] plane changes dramatically with spatial position over the Galaxy \citep{Nidever2014,Hayden2015}.
However, as pointed out in \cite{Bland-Hawthorn2019}, a star's kinematic changes throughout its lifetime but not its chemistry. As a result, if it is advantageous to break up the disk into constituents to study it, it is often desirable to divide it in the chemical rather than the dynamical plane. 
The high and low-$\alpha$ disks have different element abundance ratios in a multitude of elements, indicative of their different star formation histories \citep[\eg][]{Bensby2014,Masseron2015}.
Recent work leveraging large data shows that the high and low-$\alpha$ sequence appear to have different dynamical properties, at all ages \citep{Mackereth2019,Gandhi2019}.

Different hypothesis have been proposed for the formation of the $\alpha$-bimodality. Vertical disk heating driven by an encounter between the Milky Way and satellite galaxies \cite{Quinn1993} and the accretion of satellite stars \cite{Abadi2003} has been invoked as potential culprits. More recent simulations demonstrate other scenarios such as clumpy formation to form the \ha\ disk \citep[\eg][]{Clarke2019,Debattista2019} or gas accretion to form the \lowalpha sequence \citep[][]{Agertz2020,Buck2020}. Regardless of the mechanisms via which the high and low-$\alpha$ disk were respectively formed, and whether or not they represent a shared or separate star formation history, the empirical differences in chemical composition and dynamical properties lead us to study these two ``populations'' separately. 

In this paper, we adopt an ad-hoc separation of the high- and \lowalpha\ disk for part of our analysis. However, we go beyond this dichotomy and explore the characteristics of the disk across a grid of chemical cells in the [Fe/H]-[$\alpha/Fe]$ plane (see section \ref{sec:results}). This is perhaps are far more powerful approach to study the disk. Indeed it is now readily enabled with large samples of stars from surveys like APOGEE. A similar line of analysis  was already suggested by \citep{Bovy2011}. Under this decomposition approach they reported a  a continuous and monotonic distribution of disk thicknesses rather than a bi-modal disk. Nevertheless, the visual appearance of an high-$\alpha$/low-$\alpha$ bimodality in [Fe/H]-[$\alpha$/Fe] space is a prediction of several models. In this sense, the bimodality is broadly indicative of different formation mechanisms of the two populations, even if an exact, simple division between the apparent populations in chemical space may be undesirable for a number of analyses.

We first examine the global properties of the high- and low-$\alpha$ disk.
We investigate how their mean age distributions change spatially and chemically (how the mean age distribution changes at different location in the \alphafe-[Fe/H] plane).
We then investigate the overall age-metallicity relation (where by metallicity we refer to [Fe/H]), which can place broad constraints on galactic and chemical evolution of the Galaxy \citep[\eg][]{Edvardsson1993, Casagrande2011}.
Studies have revealed a large range of stellar ages at any fixed metallicity throughout the disk \citep[\eg][]{Feuillet2019,jonsson2020}, showcasing that [Fe/H] itself is not a chemical clock. 
Finally, we quantify the relationship between ages and individual chemical element abundances using red clump stars. We identify the red clump stars in the APOGEE catalogue from their spectra, using data-driven modeling of the correlation between flux variability and evoluionary state \citep{Hawkins2018}. Red clump stars provide a narrow region in evolutionary state, thus mitigating systematic imprints of abundance variation in the data \citep{Jofre2019}. Furthermore, they enable precision distance estimates across a large radial extent. 

Specifically, we examine the age-abundance properties for 16 elements for stars of the low- compared to the high$-\alpha$ disk. Studies of detailed age-individual chemical abundance relations - at fixed metallicity - have previously found low intrinsic dispersion for stars around these relations  \citep[\eg][]{Ness2019,Bedell2018, Sharma2020, Hayden2020}. However, these studies have focused on all the stars in the disk or only stars in the low-$\alpha$ disk. The small intrinsic dispersion around the individual age-abundance relations ($\approx$ 0.02 dex on average for the low$-\alpha$ disk) implies we can use these element abundance trends to age-date stars. 

The age-individual chemical abundance results set out strong constraints on nucleosynthetic channels and the initial composition of the star-forming gas. Examining these \textit{separately} for the high and low-$\alpha$ disk gives us insight as to which properties are shared and which are distinct across this chemical plane. This gets toward understanding the relationship between these sequences and if the high-$\alpha$ disk could be an ancestor of the low, or if its formation channel must be entirely distinct. In detail, we compare and contrast the age-abundance relations and the intrinsic dispersions around these relations. We highlight which elements are most similar between the two disks and which are least similar. We examine the mean age distributions, both spatially and in the chemical-plane, and showcase the signatures of radial migration. Using the age variable in concert with metallicity directly demonstrates how the formation and evolutionary signatures of the disk are imprinted in the data. We also provide reader an age catalog for 64,317 stars from APOGEE DR16 \citep{DR162020} with an mean age error of 0.25 dex (APO-CAN stars) as well as a red clump catalog with 22,031 stars with a contamination rate of 2.7\%.  

Section \ref{subsection:data} describes the data used in this project. Section \ref{subsection:method} details how we determined ages and red clump membership, and how we separated the high- and \lowalpha disk. 
In Section \ref{subsection:global}, we look at the overall age distribution of the APO-CAN stars from APOGEE DR16 across the Galaxy.
In Section \ref{subsection:subglobal}, we investigate the age distribution of the stars in the chemical plane at different locations in the Galaxy. Then, we examine the temporal and spatial distributions of the high- and \lowalpha disk in a grid of chemical cells across [$\alpha$/Fe]-[Fe/H].
In Section \ref{subsection:trends}, we explore the detailed age-element abundance trends for 16 different elements and the age-metallicity relation for the high- and \lowalpha disk.
Finally, in Section \ref{sec:dis}, we compare our results to  simulations.

\section{Data \& Methods} \label{sec:datamethod}
We used two data sets for this work -- the APOGEE survey DR16 spectra and abundances data \citep{DR162020}, and the APOKASC catalogue that contains ages and asteroseismic parameters \citep{Pinsonneault2018}.
The APOGEE spectrograph has a resolution of $R = 22,500$ and is mounted on the 2.5-m telescope of the Sloan Digital Sky Survey \citep{wilson+2019,gunn+2006}. For details on the data reduction process, see \cite{nidever+2015}. APOGEE spectroscopic analysis is performed using the APOGEE Stellar Parameter and Chemical Abundance Pipeline \citep[ASPCAP;][]{garcia_perez+2016}, with temperatures calibrated to the infrared flux method scale of \cite{ghb09} \citep{holtzman+2015}.

\begin{figure}[htp]
    \centering
    \includegraphics[width=0.4\textwidth]{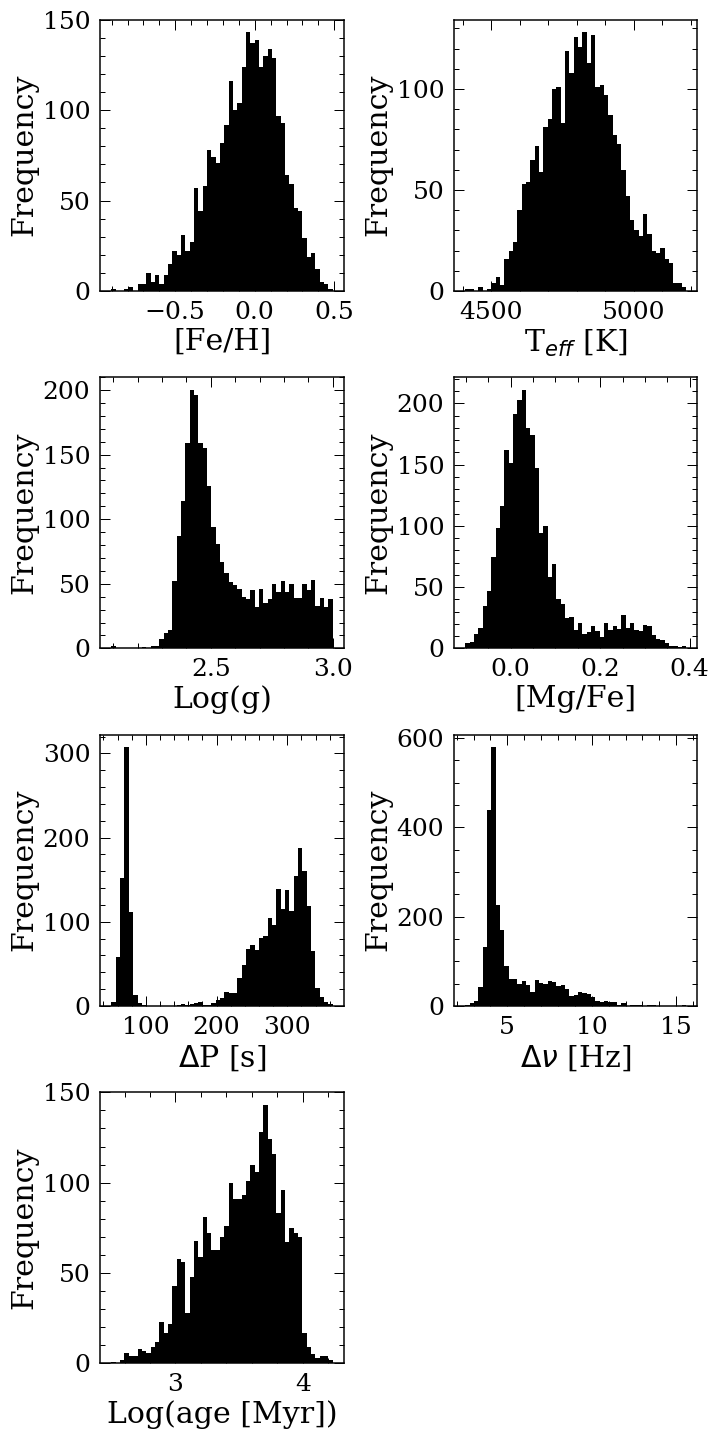}
    \caption{Histograms of the parameters in our training set of 2,616 stars for \texttt{The Cannon}.}
    \label{fig:training}
\end{figure}

We combined both data sets to examine the abundance-age relations.
The APOKASC catalogue serves not only as a benchmark data set of stars with precision ages, but also as a training set for the data driven approach of \texttt{The Cannon} to estimate ages for the rest of the red giant stars in APOGEE from their spectra and to identify red clump stars.

We worked in a narrow region of \teff-\logg\ for our analysis to circumvent systematics \citep[\eg][]{Jofre2019} and we selected the red clump stars for our endeavour as we can determine precise distances for them to use in our follow on dynamical analyses of these stars and their age-abundance relations. 

In order to create the age and the red clump catalog, we used \texttt{The Cannon} \citep{Ness2015}\footnote{Available at https://annayqho.github.io/TheCannon/intro.html}.
\texttt{The Cannon} is a data driven approach to derive stellar parameters from stellar spectra. 
Here we use a quadratic combination of the labels to predict each pixel of the spectrum, as is consisent with previous implementations \citep[\eg][]{Ness2015,Ho2017,Casey2017,Wheeler2020}. 

\subsection{Data}\label{subsection:data}
In order to use \texttt{The Cannon} to determine stellar ages and identify the red clump stars, we needed a high fidelity set of reference objects for training the model. \texttt{The Cannon} is a tool that determines the relationship between flux and labels that describe the variability of the flux. Therefore, it is important to include the labels that describe most of the variability in the flux, hence we included the set of labels of metallicity, \teff, \logg, and [Mg/Fe] in addition to the asteroseismic parameter and ages that we wished to infer for APOGEE DR16 spectra.  

We used measurements of frequency spacing between $p$-modes, \dnu, and period spacing of the mixed $g$ and $p$ modes, \dP, from \cite{Vrard2016} for 6,111 \kepler\ stars. We obtained estimates of ages  from the second APOKASC catalog \citep{Pinsonneault2018} and parameters of \teff, \logg, [Fe/H], and [Mg/Fe] from APOGEE's DR16 data release \citep[][]{jonsson2020}. 
We included \dnu\ and \dP\ since these astroseismic parameters can be used to better separate the red clump stars with the red giant branch stars \citep{Bedding2011,Ting2018}.
After cross-matching these two catalogs, we were able to find 2,616 common stars to construct the training set. Figure~\ref{fig:training} shows the parameter space occupied by the training set.

The distances that we use in our analysis are from \texttt{StarHorse} \citep{Queiroz2018}, which is a bayesian tool for determining stellar masses, ages, distances, and extinctions for field stars.
To study the detailed age-element abundance relations, we also included 16 individual element abundances, C, N, O, Mg, Al, Si, S,K, Ca, Ti, V, Mn, Ni, P, Cr, Co from the APOGEE DR16 catalog, inferred using ASPCAP \footnote{Available at \url{https://www.sdss.org/dr16/irspec/spectro_data/}.}. 
We removed Na as this showed anomalous behaviour and indeed in \cite{Ness2019} was unable to be recovered in cross-validation with The Cannon.

We downloaded APOGEE DR16 spectra  from the SDSS-IV Science Archive Server (SAS)\footnote{Avaliable at https://data.sdss.org/sas/}.

\subsection{Methods}\label{subsection:method}
\subsubsection{Creating the age/red clump catalog with \texttt{The Cannon}}

For our implementation of \texttt{The Cannon} we use a second-order polynomial to fit the spectra flux ($F$) for each star, $n$, with labels at each wavelength, $\lambda$.
The labels for each star used in this project, in vector form, is $l_n$ = [\teff, \logg, [Mg/Fe], [Fe/H], \dP, \dnu, and $Log_{10}$(age)].
As a result, the model can be described as:
\begin{equation} \notag
\begin{split}
    F_{n\lambda}&=\theta^0_\lambda \\
    &+\theta^{T_{\mbox{\scriptsize eff}}}_\lambda T_{\mbox{\scriptsize eff}} + ... + \theta^{\log_{10}(age)}_\lambda \log_{10}(age) \\
    &+\theta^{T_{\mbox{\scriptsize eff}}^2}_\lambda T_{\mbox{\scriptsize eff}}^2 + ... + \theta^{\log_{10}(age)^2}_\lambda \log_{10}(age)^2\\
    &+\theta^{T_{\mbox{\scriptsize eff}}\log g}_\lambda T_{\mbox{\scriptsize eff}}\log g + ... + \theta^{\log_{10}(age)\Delta \nu}_\lambda \log_{10}(age)\Delta \nu \\
    &+error
\end{split}
\end{equation}

In order to infer the stellar parameters, we have to train \texttt{The Cannon} on a set of reference stars in order to fit for the coefficients, {$\theta_\lambda$}.
To train \texttt{The Cannon}, we first excluded stars that were flagged as ``bad'' (stars with \textit{ASPCAPFLAG} flag 23), and/or had signal-to-noise ratio less than 100.
This left us with 2,480 stars with 7 labels in our training set of stars (parameter range shown in Figure~\ref{fig:training}). 

\begin{figure}[htp]
    \centering
    \includegraphics[width=0.5\textwidth]{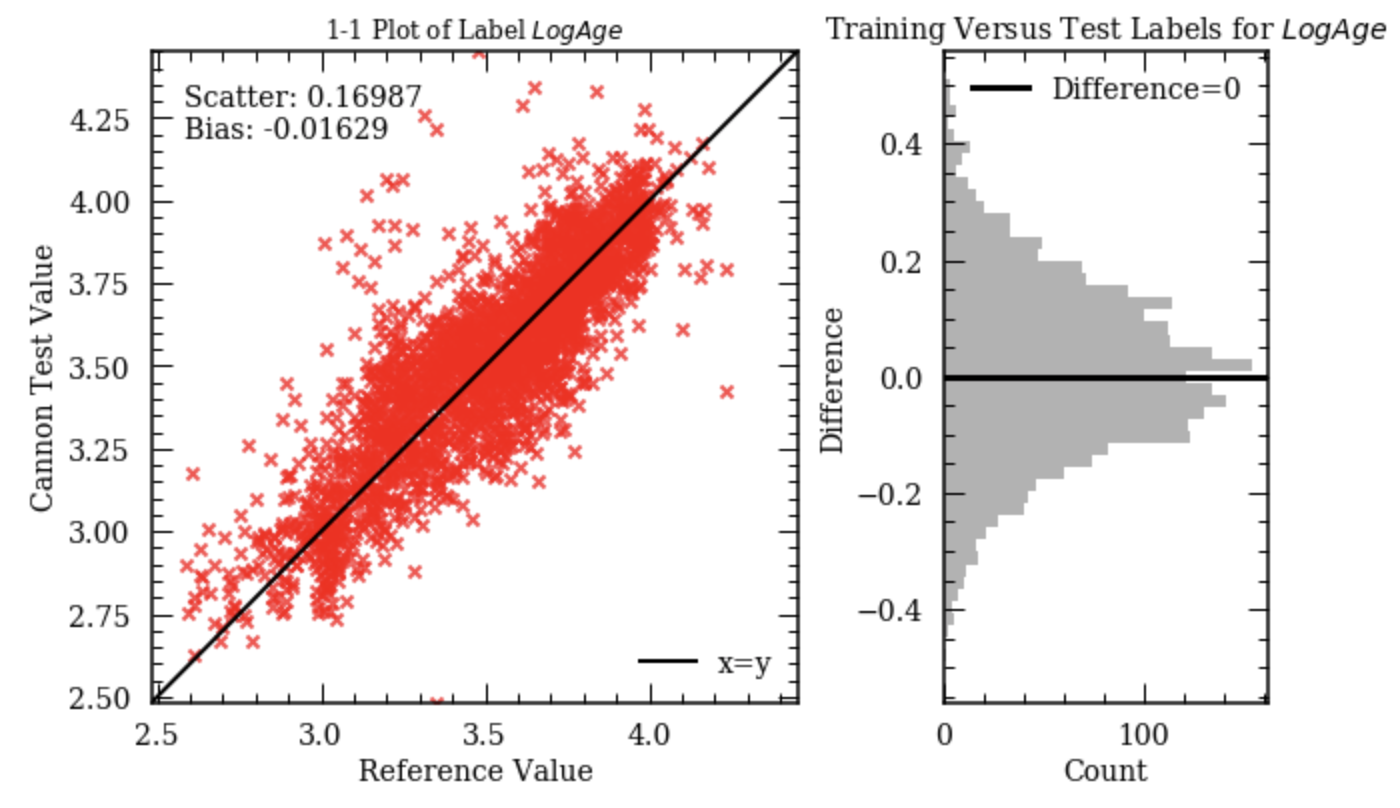}
    \caption{10-fold cross-validation result for stellar ages estimated with \texttt{The Cannon} for all 2,480 stars. 
    The ages are in unit of Myr.}
    \label{fig:cvresult}
\end{figure}

To test the performance of \texttt{The Cannon}, we performed a 10-fold cross-validation test, in which we left 10\% of the data untouched and trained the model on the rest of the 90\% and predicted the labels for stars in the rest of that 10\%. 
We then repeated the same test 10 times with a different 10\% of the data left out (hence called 10-fold).
The cross-validation result for stellar age is shown in Figure~\ref{fig:cvresult}.
We added the cross-validation root mean squared (rms) scatter to the error estimated from \texttt{The Cannon} to obtain the final systematic age uncertainty, which yield a median uncertainty of around 1.9 Gyr across all ages.
The rms scatter for other labels is --- 0.017 dex for [Fe/H], 26.8 K for \teff, 0.056 dex for \logg, 0.028 dex for [Mg/Fe], 40.1 s for \dP, and 0.6 $\mu$Hz for \dnu.
After training \texttt{The Cannon}, we applied the trained model to the rest of the APOGEE DR16 spectra. 
One caveat of using data-driven method is that we were not able to (reliably) infer stellar parameters for stars outside of the range of those values of the training set   since the \texttt{The Cannon} extrapolates beyond the training sample regime.
Therefore, we first discarded stars outside of the training parameter range.
We only included stars with \teff\ between 4,400 K - 5,200 K, \logg\ between 2.2 dex - 3.5 dex, and [Fe/H] between -0.8 dex - 0.5 dex. 
We also excluded stars with abnormal element abundances (absolute abundance values $>$ 1 dex) for the 16 elements we are interested in.
This left us with 64,317 stars.

\begin{figure}[htp]
    \centering
    \includegraphics[width=0.5\textwidth]{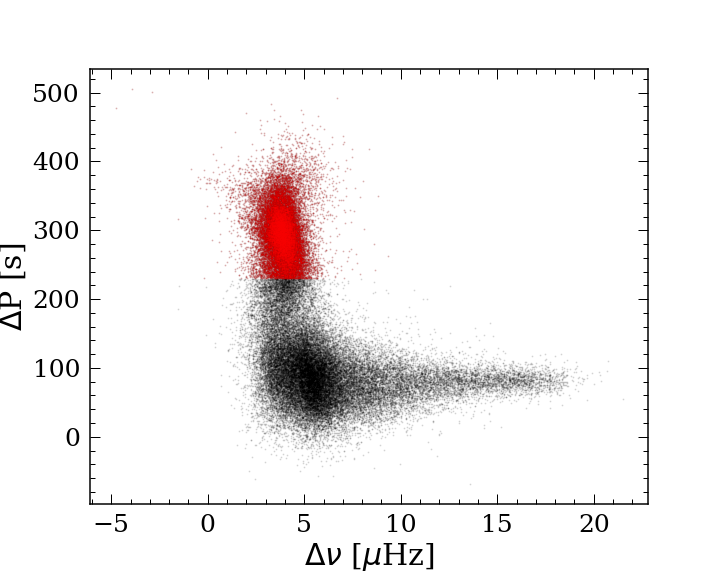}
    \caption{\dP\ vs \dnu\ for the APO-CAN stars (black).
    The red dots show the 22,031 red clump stars (\dP\ $>$ 230 s) in this parameter space.
    It is clear that the \dP\ separates the red clump stars and the red giant branch stars.}
    \label{fig:RCdet}
\end{figure}

To create the red clump catalog, we follow the method described in \cite{Ting2018} to select stars with \dP\ $>$ 230 s as red clump stars.
Figure~\ref{fig:RCdet} shows our results. 
The black dots show all the stars in \dP-\dnu\ space and the red clump stars are shown in red. 
It is clear that \dP\ separates the two types of stars.
We calculate the contamination rate following their method, where we measured the false positive rate by taking the ratio of stars that have predicted \dP\ $>$ 230 s but true \dP\ $<230$ s.
This yield a contamination rate of 2.7\%, which is extremely similar to that from \cite{Ting2018}.
However, we might be underestimating the contamination rate with this method since by assuming their catalog is the ground truth, we misclassify 13\% of the stars.

\subsubsection{Separation of the high- and \lowalpha disk}

For the purpose of examining the different properties of the high and low$-\alpha$ stars, we separated the high- and \lowalpha disk with an ad-hoc line, (0.1$\times$[Fe/H]+0.063)  
We also explored separating the two disks with a clustering algorithm described in \cite{Ratcliffe2020} in the Appendix. 
Figure~\ref{fig:alphadisks} shows the APO-CAN stars in the [$\alpha$/Fe]-[Fe/H] plane.
The left plot shows the stars colored by age and the right plot shows the high- (red) and the \lowalpha (blue) disk.
This color code will be used throughout the paper to distinguish between these two disks.
It is clear that the stars in the \ha\ disk are on average older than those in the \lowalpha disk, which is what we expected. 
Within the \lowalpha disk, stellar ages increase with [$\alpha$/Fe].

\begin{figure*}[htp]
    \centering
    \includegraphics[width=\textwidth]{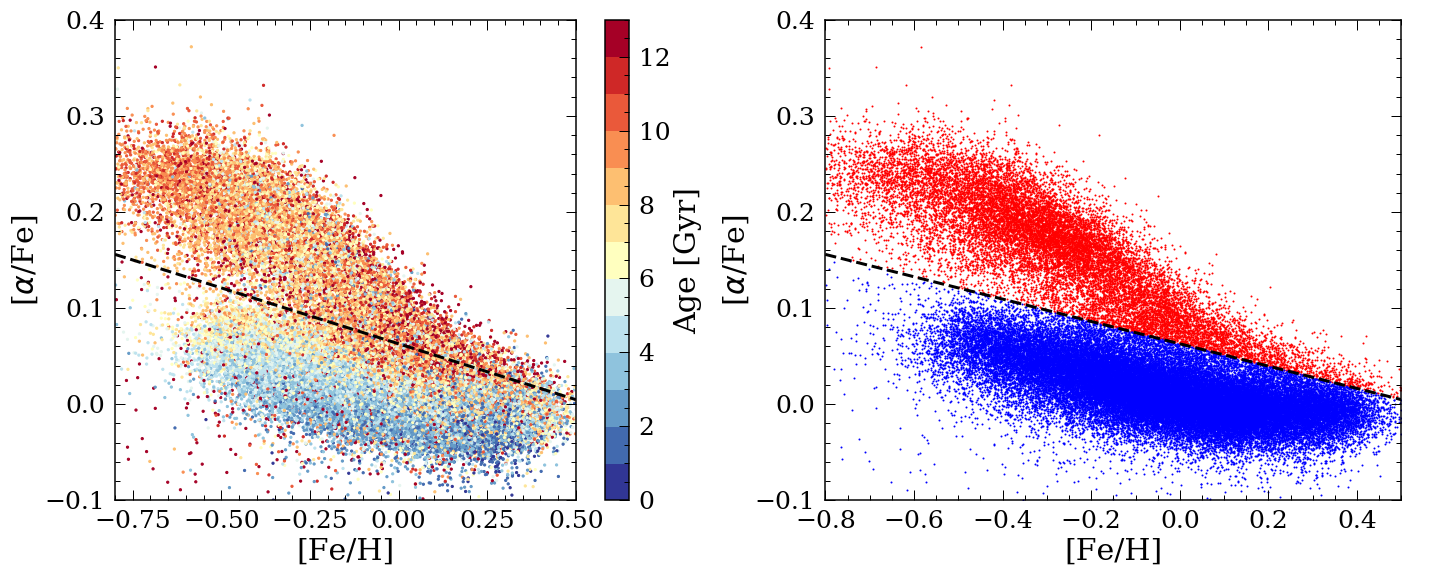}
    \caption{64,317 APO-CAN stars plotted in [$\alpha$/Fe]-[Fe/H] space. 
    The black dotted line separates the high- (red) and \lowalpha (blue) disk.
    The bi-modal distribution of these two disks and the positive $\alpha$-age gradient are clearly visible. }
    \label{fig:alphadisks}
\end{figure*}

\section{Results}\label{sec:results}
In this section, we first examine the global age and metallicity trends across the disk.
We then explore the age distribution in the chemical plane; first,  across the disk spatially, and then across small cells in  [$\alpha$/Fe]-[Fe/H] (section \ref{subsection:subglobal}).
In section \ref{subsection:trends}, we report the age-chemical abundance trends for 16 elements and calculate the intrinsic dispersions around these relations. We examine the high- and \lowalpha disk separately, comparing and contrasting the relations and their intrinsic dispersions.

\subsection{The global age/metallicity skew distribution across the Milky Way: two episodes of star formation}\label{subsection:global}

\begin{figure*}[htp]
    \centering
    \includegraphics[width=\textwidth]{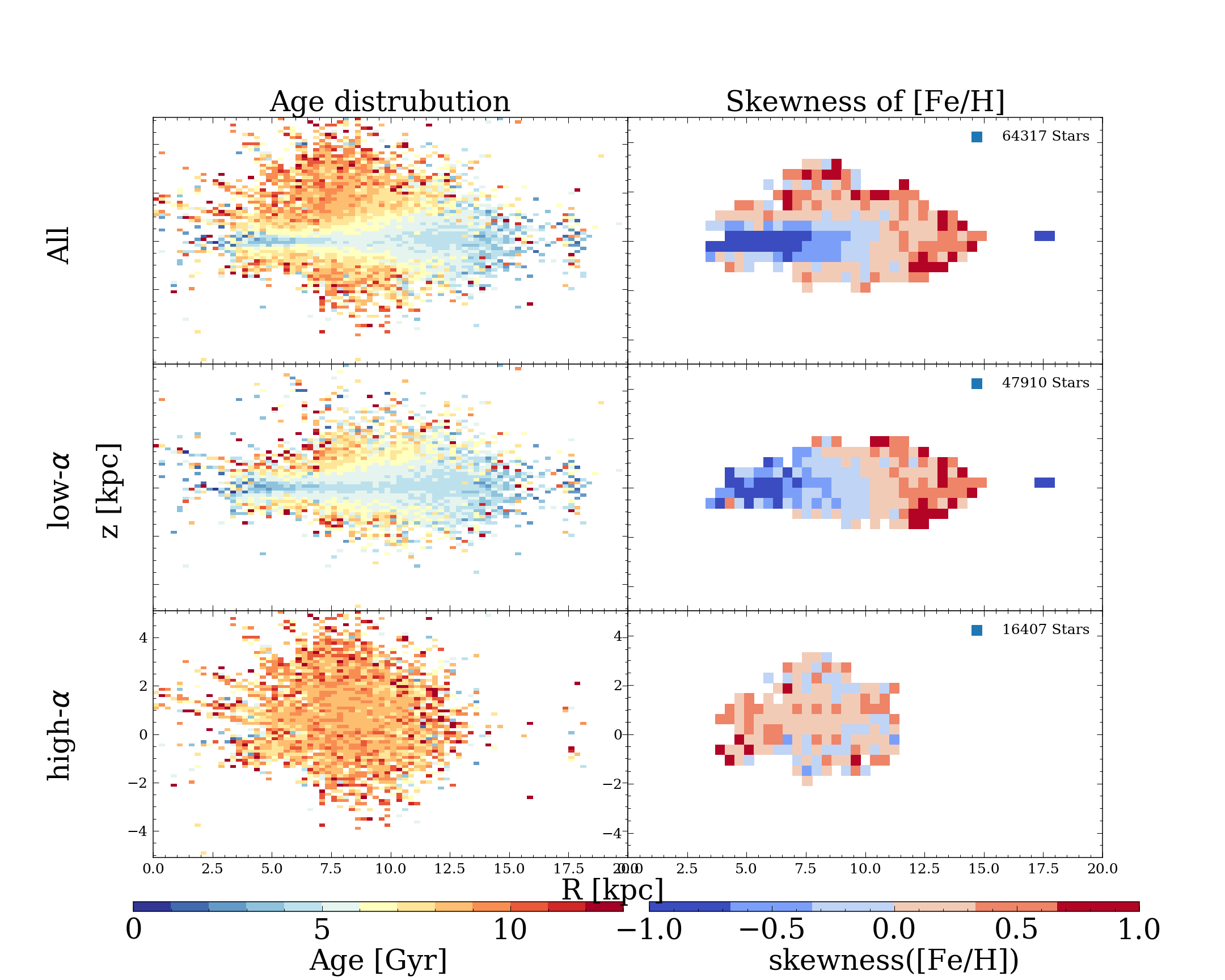}
    \caption{
    The top plots show the overall age distribution/metallcity skewness, the middle plots shows the age distribution/metallcity skewness of the \lowalpha disk, and the bottom plots shows the age distribution/metallcity skewness of the \ha\ disk. 
    The overall age distribution of the stars and flaring of the \lowalpha disk stars suggests an inside-out formation history and the negative skew in metallicity suggests radial migration (for more explanation, see section \ref{subsection:global}).
    The lack of any age gradient across the Galaxy in the \ha\ disk places constraints on the enrichment rate and formation history of the \ha\ disk. 
    }
    \label{fig:agedist}
\end{figure*}

In Figure~\ref{fig:agedist}, the top left plot shows the age distribution of the APO-CAN stars with inferred ages from \texttt{The Cannon} across the APOGEE footprint (mean age: 6.3 Gyr; standard deviation: 3.3 Gyr).
These stars range from $R$ = 0 $\sim$ 18 kpc in radius and $|z|$ $<$  5 kpc in galactic height from the plane.
It is worth pointing out that this range does not reflect the underlying density distribution of the disk, but only the observing strategy and selection function of APOGEE.
However, even without correcting for the selection function, we are still able to see clear mean trends across the disk, which are signatures of its formation.
The middle left plot shows the mean ages of the \lowalpha disk stars (45,983 stars; mean age: 5.4 Gyr; standard deviation: 2.9 Gyr); and the bottom left plot shows the mean age distribution of the \ha\ disk (16,416 stars; mean age: 8.8 Gyr; standard deviation: 3.1 Gyr).
We found that 7\% of the $\alpha$-enriched stars are younger than 5 Gyr, which are also observed by \cite{Martig2015,Chiappini2015,Feuillet2018}. 
Within the \lowalpha disk stars, 15\% are older than 8 Gyr.

Looking first at the age distribution of the \lowalpha disk: Young stars are concentrated to the mid-plane and stars with larger ages are seen higher above the mid-plane, as expected \citep[see also][]{Ness2016, Mackereth2017}.
Looking at the middle left figure, the \lowalpha disk shows flaring in the young population \citep{Mackereth2018, Bovy2016} and at a given height from the plane, $|z|$, the mean age decreases with radius, R.
The concentration of old stars in the inner region and young in the outer is indicative of inside-out formation of the Milky Way disk.
Looking, second, at the age distribution of the \highalpha\ disk: There are no age gradients across either R or $|z|$.
The \highalpha\ disk stars are old even along the mid-plane and extend to larger heights compared to the \lowalpha stars.

There are a range of ages in the \highalpha\ disk. Yet, the absence of any age gradient suggests a rapid formation history for the \highalpha\ disk where all the stars were formed early, before the stars of the \lowalpha disk.

Next, we examine the spatial distribution of the metallicity skew across the disk, where the skewness is used as a measure of how far the distribution deviates from a Gaussian. 
A positive age skew means there is an of stars with ages older than the average stellar age in that bin, and visa-versa.  
The top panel shows the skew in all stars, the middle in the \lowalpha disk and the bottom for the \highalpha disk.

\cite{Hayden2015} and also \cite{Loebman2016} highlighted the change in the direction of the metallicity skewness across radius in the disk as a possible signature of radial migration. 
In the presence of migration and an initial disk metallicity gradient, the distribution can skew in opposite ways in the inner and outer region, respectively, as stars migrate in and out, across the disk.  

We calculated the skewness of the metallicity distribution in spatial bins with a bin size of (R, z) = (0.4 kpc, 0.4 kpc) using the \texttt{Python} package \texttt{scipy.stats.skew}, excluding any bins with fewer than 20 stars.

From the middle right plot, we can see that there is a clear trend from negative skewness to positive skewness as we move from the inner Galactic disk to the outer disk in the \lowalpha disk  \citep[see also][]{Kochukhov2021}.
This supports the idea that the disk formed with a negative metallicity gradient, and radial migration has been significant in the \lowalpha disk.

We note there is a strong negative skewness for the group of stars around R = 17.5 kpc ($\sim$ 140 stars).
We further investigated these stars and we did not find any specific APOGEE programs associated with these stars.
This group of (on average) young stars has an excess number of metal poor stars.
We note that the mean metallicity of these stars is on average higher than that of stars around R= 12 kpc and is in fact similar to that of stars that are around R= 5 kpc.

For the \ha\ disk (bottom right plot), there is a weak positive metallicity skew across (R,z), but no gradient in the skew across R as seen for the \lowalpha disk above it.  However, this does not indicate that radial migration is not significant. Rather, this is consistent with the \ha\ disk having no \textit{initial} metallicity gradient, at formation; subsequently any migration would not affect the metallicity skewness across radius, in each spatial bin. 
The lack of metallicity gradient in the \ha\ disk is also seen in cosmological simulations \citep[e.g.][]{Agertz2020}.
In the \ha\ disk there is an overall positive skewness on the order of 0.19 dex in metallicity.
Presumably this places constraints on the enrichment rate and formation history of the \ha\ disk.

\subsection{The age distribution of the chemical plane across the Galaxy: signatures of radial migration and two modes of star formation} \label{subsection:subglobal}

\begin{figure*}[htp]
    \centering
    \includegraphics[width=\textwidth]{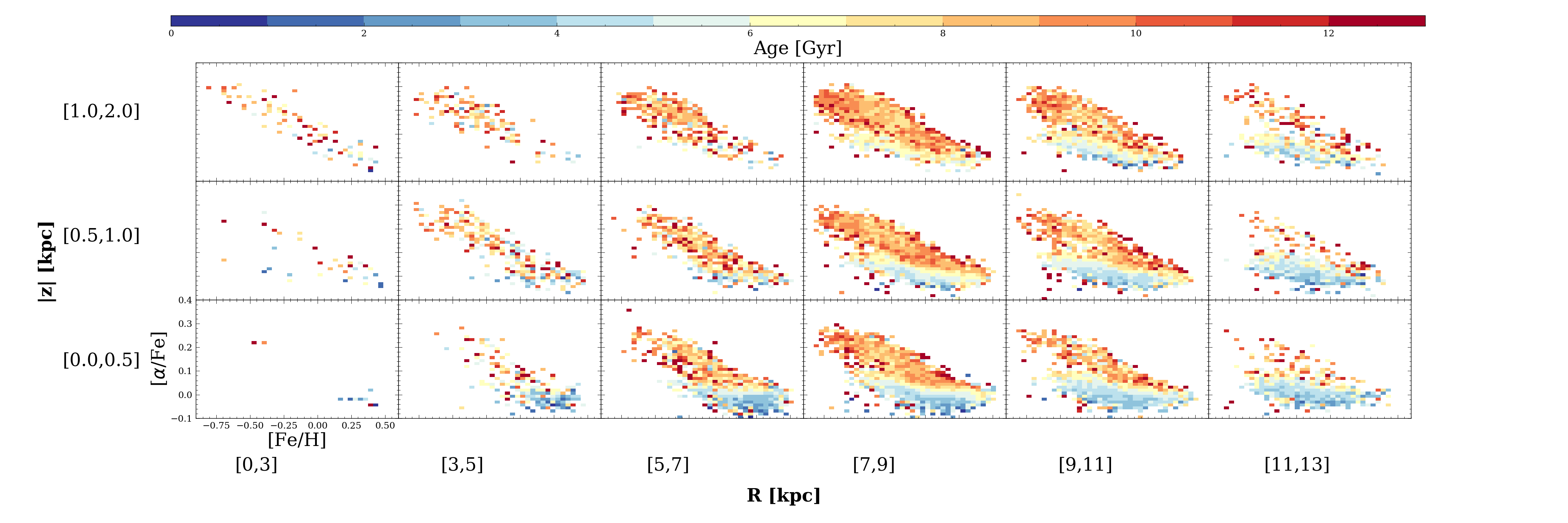}
    \caption{The age distribution of the APO-CAN stars in APOGEE DR16 in mono spatial bins inspired by Figure~4 in \cite{Hayden2015}.
    The  radius increases moving to the right, and the Galactic height increases moving towards the top.
    The \ha\ disk dominates as the Galactic height increases, and the \lowalpha\ disk dominates at larger Galactic radius. 
    The \lowalpha\ stars at large Galactic height in the outer disk is the signature of the flaring young disk. 
    }
    \label{fig:age_space}
\end{figure*}

\begin{figure*}[htp]
    \centering
    \includegraphics[width=\textwidth]{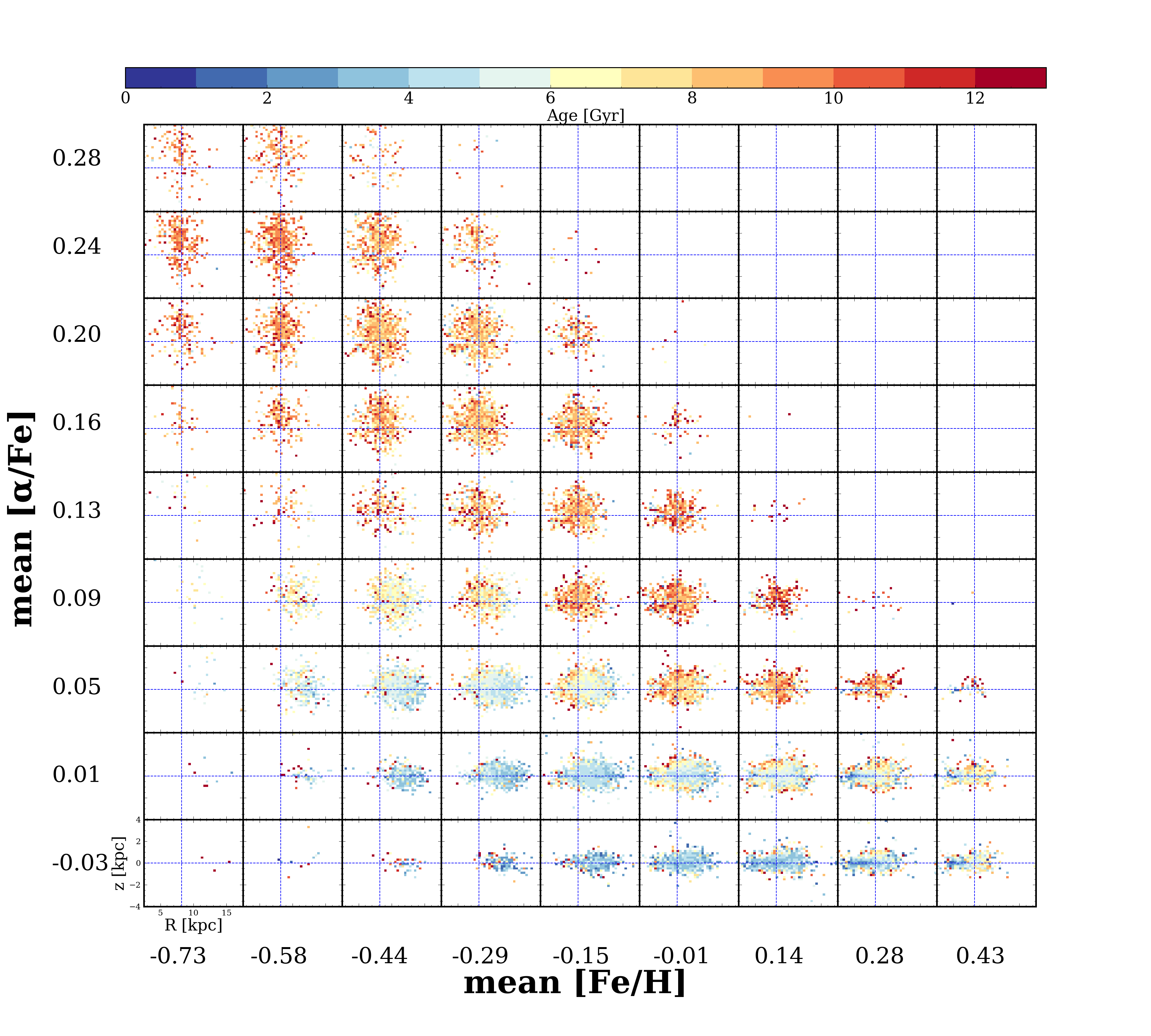}
    \caption{Spatial age distribution for all APO-CAN stars in mono \alphafe-[Fe/H] bins. 
    The mean values of \alphafe\ and metallcity for each subplot are displayed along the axis, and the bin sizes are 0.039 dex and 0.14 for \alphafe\ and metallcity, respectively. 
    The metallicity increases towards the right, and stars are more $\alpha$ enhanced towards the top.
    The blue dashed line mark the $z$=0 and solar radius.
    Every column is at a fixed [Fe/H], and every row is at a fixed \alphafe.
    The transition from low- to \ha\ disk as the $\alpha$ element increases is gradual in the chemical cells, which means in order to understand the formation of these two disks in the chemical plane, one should consider a smooth transition function instead of separating them into two disks. 
    The skew and dispersion of ages in each bin are shown in Figure~\ref{fig:age_chem_skew} and Figure~\ref{fig:age_chem_disp}.
    }
    \label{fig:age_alpha}
\end{figure*}

We dissect the APO-CAN stars into spatial (figure~\ref{fig:age_space}) and chemical (figure~\ref{fig:age_alpha}) bins in order to piece-wise examine the structure of these two disks.

Figure~\ref{fig:age_space} shows the age distribution of stars in [$\alpha$/Fe]-[Fe/H] plane at different locations of the Galaxy's disk. From left to right, the Galactic radius increases, from $R$ = 0 to 13 kpc, and from bottom to top, the absolute vertical height increases, from $|z|$ = 0 to 2 kpc, similar to the figure in \cite{Nidever2014,Hayden2015} but coloured here by age.
Moving away from the galactic plane, the mean stellar age clearly increases, and in the inner disk at small Galactic radius, the (older) \ha\ disk stars dominate. 
On the other hand, the (younger) \lowalpha\ disk dominates nearer to the mid- plane and in the outer Galactic disk, where \lowalpha\ stars with large galactic heights in the outer disk indicate signatures of flaring.
It is clear that a significant mean age gradient is imputed across (R,z) merely by the changing ratio of the number of (younger) \lowalpha\ and (older) \highalpha\ stars, respectively.

We qualitatively compare this result to the expectation from simulations (Figure~A1 and B1 top two plots) from \cite{Buck2020}. These simulations suggest that the $\alpha$-bimodality is a generic consequence of a low-$\alpha$ disk forming gas-rich merger after the high-$\alpha$ disk is in place. The simulations presented in \citet{Buck2020} reveal a similar age gradient along the [$\alpha$/Fe] axis in the \lowalpha disk at any spatial bin whilst almost no age gradient is found for the \highalpha\ disk. Overall, in both simulation and data, there is a strong age gradient with [$\alpha$/Fe] (at fixed [Fe/H]) and a  weak gradient with [Fe/H] (at fixed [$\alpha$/Fe]. The shape of the distribution of stars with similar ages in each spatial bin in Figure 6 is inclined with respect to the [Fe/H] axis. 
Interestingly, the inclination in the age distribution of the low-$\alpha$ disk as shown across (R,z) in this figure is only seen in the simulation with a strong bar \citep[Fig.~A1 in][]{Buck2020}.


The simulations further show a similar trend of increasing [Fe/H] with decreasing radius. Note, this comparison is only of qualitative nature since no APOGEE selection function nor age uncertainties were taken into account in their work.

From Figure \ref{fig:age_space}, we see that there is no age gradient for the \ha\ disk along either [Fe/H] or [$\alpha$/Fe], at any given location that has been observed. The absence of any age gradient in the \ha\ disk across [Fe/H] or [$\alpha$/Fe] is consistent with the results from \cite{Agertz2020}, in which, similarly to \cite{Buck2020} they connect the last gas rich merger as a \lowalpha\ disk formation mechanism  around the in-place \ha\ disk \citep{Renaud2020}

We have examined the mean spatial trends of age across the overall chemical, \alphafe-[Fe/H], plane. We now look deeper into the conditions of the star-forming gas by examining the age distribution across small cells in \alphafe-[Fe/H] across the disk. This is related to the analysis route of \citet{Bovy2011}, who examined the scale height and lengths of mono-abundance populations.

Figure~\ref{fig:age_alpha} shows the age distribution of the same APO-CAN stars as shown in Figure~\ref{fig:age_space}, in a grid of \alphafe-[Fe/H] bins.
We call these chemical cells (they are not strictly `mono' abundance populations; the cells are larger than the errors on the [Fe/H] and [$\alpha$/Fe]). The metallicity, [Fe/H] increases toward the right and \alphafe\ increases upwards. 
The bin sizes are 0.04 dex in \alphafe\ and 0.14 dex in [Fe/H].
Each individual cell shows the spatial distribution of stars in the $R$-$z$ plane.
The blue dashed lines show the $z$=0 plane and the radius location of the sun. 

Across the matrix of chemical cells, we see different portions of the stellar disk, that together, comprise the full disk stellar distribution. Globally, the chemical cell projection shows a chemical population of spatially and temporally distinct disks,  that looks to be consistent with an inside-out and upside-down formation process \citep{Bird2013}. Note, we do not take into account the selection function. 
Nonetheless, the age gradient of older to younger stars from the inner to outer region and spatial flattening across cells of mean decreasing age is indicative of these processes. 

Figure 7 shows that the transition from the young to old stars is marked and rapid across the chemical plane, in that the transformation happens within a small range of [Fe/H]-\alphafe, across chemical cells. Outside of these clear and most dramatic changes across chemical cells, there are subtle spatial and temporal variations along both rows of [Fe/H] and columns of [$\alpha$/Fe]. We see using the chemical cells, that away from where there are strong age gradients across cells, most cells do not show age gradients within them.

Along fixed rows in ([Fe/H]) the disks are far more similar than across columns of [$\alpha$/Fe]. Most notably, moving along rows reveals subtle age gradients and shifting radial distribution of the stars. Along fixed columns in [$\alpha$/Fe], there are strong changes in the mean age of populations and more dramatic spatial changes. Looking along say the forth column from left, the bottom row shows young stars concentrated to the outer disk and distributed around the mid-plane, and old stars concentrated to the inner Galaxy and diffusely distributed around the plane at top.

The matrix of the chemical cells reveals an inversion of the age gradient along [Fe/H] moving from highest to lowest [$\alpha$/Fe]. 
Looking along the top five rows, for the chemical cells with highest $\alpha$-enhancement, there is a mean decrease in age as [Fe/H] increases.
Correspondingly, along these rows, the stars are less diffusely distributed around the plane spatially, and as metallicity increases the mean radius of the stars increases.  
Note that in chemical cell projection, along the bottom rows, the age gradient inverts; the stars become older moving to increasing [Fe/H]. 
However, they also are on average nearer to the Galactic center, which is the reverse spatial trend to high [$\alpha$/Fe] stars. 
This is indicative of the initial negative [Fe/H] gradient in the low-alpha disk. 
The flattening distribution of stars around the mid-plane as they become younger across the chemical cells, and at fixed [$\alpha$/Fe], is indicative of an upside-down formation of an ensemble of disks represented within the chemical cells. 
The oldest stars in the disk at the high [Fe/H] and low [$\alpha$/Fe] have presumably migrated from the inner disk, which would explain this age gradient inversion and change in spatial trend compared to the high-$\alpha$ rows of stars.

We repeated this analysis using the ages included in the APOGEE DR16 release, derived using a neural network \citep{Mackereth2019}. We found a small but important difference in doing this:  the most metal-poor \ha\ stars are \textit{younger} than the more metal-rich \ha\ stars using the ages provided in \citet{Mackereth2019}. This is the opposite to what we find using our age catalogue and this instead suggests a outside-in formation of the \ha\ disk. 
The differences in the results from our inference using The Cannon and the neural network approach are likely a consequence of sparse training data in this chemical realm. This small discrepancy has significant implications for the formation history inferred under the different catalogues for the high-$\alpha$ stars. 

We also investigated the skew and standard deviation of ages in this plane and found the \ha\ disk has the strongest negative age skew, and the \lowalpha disk has the strongest positive age skew (see Figure~\ref{fig:age_chem_skew} in Appendix). 
The largest age dispersion is seen for the low $\alpha$ and metal rich stars (see Figure~\ref{fig:age_chem_disp} in Appendix). 
This large dispersion could support the two in-fall analytic model described in \cite{Spitoni2021}. Overall, Figure 7 suggests the disk can be well considered as a continuum of populations, rather than two distinct populations.  
The nature of the changing age distribution in chemical cells resonates with the scale height and scale length analysis of \cite{Bovy2012}.

Now that we have examined the age and spatial distribution of our stars in chemical cells, we turn to the age-metallicity relations of stars and their detailed age-abundance trends for 16 individual elements. In doing so we seek to quantify the star formation environment in chemical enrichment over time across the disk.

\subsection{Age-metallicity trends in the high- and \lowalpha disk}\label{subsection:agemetal}\label{subsection:Feage}

We now explore the age-metallicity relations for our sample across its spatial extent (R, z). This is similar to the analysis of \cite{Feuillet2019}, who examined the age-metallicty relation for stars in their sample at different Galactic radius and heights (Figure~3 in their paper). They subsequently argued the trends to be signature of radial migration \citep[see][]{Minchev2013, Buck2020}. We employ this same analysis but we separate the stars into high and low-$\alpha$ populations. Since high and low-$\alpha$ stars have different initial conditions, spatial and temporal distributions, it is a natural next step to examine the age-[Fe/H] relations conditioned on narrower regions of chemical space (including beyond our bi-modal population model here). This should place stronger empirical constraints on the formation and evolution mechanisms of the disk - and here reveals interesting differences between the high and low-$\alpha$ sequences. 

\begin{figure*}[htp]
    \centering
    \includegraphics[width=\textwidth]{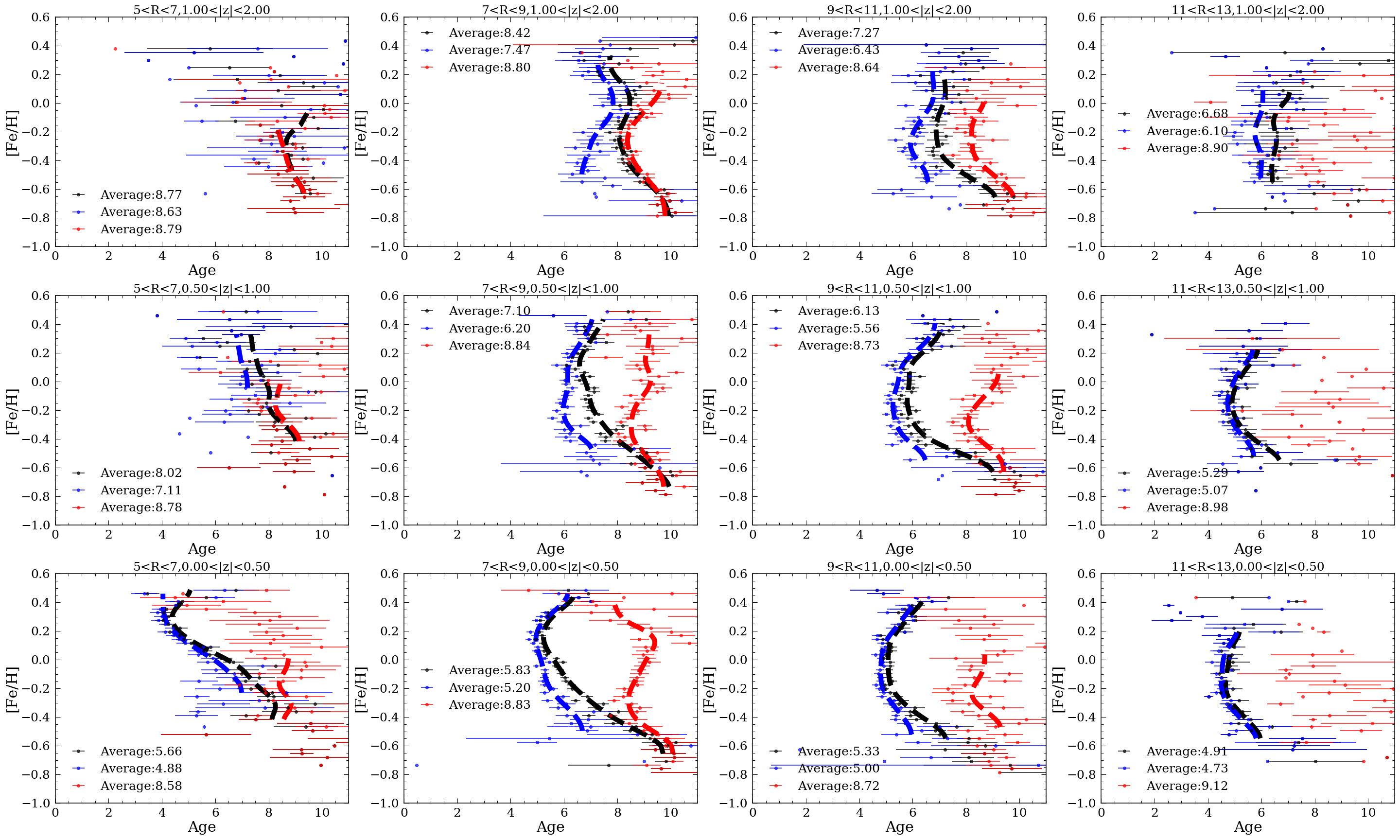}
    \caption{Age-metallicity relations from all APO-CAN stars (black), \ha\ disk stars (red), and \lowalpha disk stars (blue) in different spatial bins.
    Each point is the mean age in that metallicity bin (width of 0.027 dex), and the errorbars are calculated as $\sigma_{age}$/$\sqrt{N-1}$, where $\sigma_{age}$ is the standard deviation of the ages in each bin, and $N$ is the number of stars in that bin.
    The dashed lines are calculated using a 1D Gaussian filter (\texttt{scipy.ndimage.gaussian\_filter}, 2020SciPy) with a kernel size of 3.
    There are significant differences between the age-metallicity relations for the high- and \lowalpha disks.
    }
    \label{fig:Feage}
\end{figure*}

Figure~\ref{fig:Feage} shows the age-metallicity relation in (four) different spatial bins, separated by the high- (red) and \lowalpha (blue) disk for all the 64,339 stars. This figure spans a radial range from $R$ = 5 kpc to $R$ = 13 kpc, and 3 different Galactic height bins, from $|z|$ = 0 kpc to $|z|$ = 2 kpc.
In each spatial bin, we calculated the mean age in each small metallicity bin.
These 50 metallicity bins range from -0.8 to 0.5 dex so that each bin spans 0.027 dex.
We then calculated the standard error on the mean by $\sigma_{age}$/$\sqrt{N-1}$, in which $\sigma_{age}$ is the standard deviation of the ages in each bin, and $N$ is the number of stars in that bin, to be the uncertainty on the mean stellar ages. 
The lines connecting the mean age bins vertically are generated by running a 1D Gaussian filter with a kernel size of 3 using the \texttt{scipy.ndimage.gaussian\_filter} function from \texttt{scipy} \citep{2020SciPy}.

The black dots and dashed line show the overall trends for the entire sample.
The average age for stars in each spatial bin is shown in the legend.

Compared to Figure~3 from \cite{Feuillet2019}, in which they measured the age-metallicity trends for stars in the APOGEE DR14, our overall trends show very similar results throughout the Galaxy.
One of the main features in these relations is the primary turnover point in the age-metallicity trends, which is a presumed marker of radial migration \citep[see][]{Frankel2018, Frankel2019}. 
These turnover points locate the oldest/youngest stars in the trends where the age-metallicity relations change.
For stars with $|z| <$ 0.5 kpc, the turnover point is at $\sim$ 0.2 dex for 5 $<$ R $<$ 7. The location of this point gradually moves towards lower metallicity as the radius increases ($\sim$ 0.1 dex for 7 $<$ R $<$ 9, $\sim$ -0.2 dex for 9 $<$ R $<$ 11, and $\sim$ -0.4 dex for 11 $<$ R $<$ 13).
We found the change in the turning point metallicity seen in \citet{Feuillet2019} comes from the behaviour of the \lowalpha stars.

To aide interpretation of this figure consider the following: The age-[Fe/H] relationship in the solar neighbourhood is fairly flat \citep[][e.g.]{Nordstrom2004,Ibukiyama2002}. 
Thus, [Fe/H] is not an age indicator, and presumably stars in the solar neighbourhood have a large range of initial birth radii.
Any given spatial location will comprise stars born across the disk and at different times.
Nucleosynthetic enrichment processes increase overall metallicity over time.
Assuming an initial radial metallicity gradient in the Galaxy and self-enriching disk, this means that the first born, oldest stars would have the lowest metallicities compared to the younger stars at a fixed birth radius (assuming the [Fe/H] is a birth property and stays relatively constant throughout the stars' lifetime). 
As a result, without stars migrating throughout the Galaxy, the metallicity should monotonically decrease with increasing stellar age, which is expected from temporal gas enrichment.

However, we see turning points where the age-metallicity relation has changed, away from a presumed fiducial monotonic decrease.
Under this scenario, the most metal rich old stars in the sample that are present across all radii must have migrated from the inner region where the gas was more enriched than the stars formed at the same time at larger radii in the disk.

The most striking result in this figure is the differences in the low- and \ha\ age-metallicity relations. The  \ha\ and \lowalpha disk have different mean ages, but also show different locations of the age-[Fe/H] turning points across the galactic disk.
The difference in the turning point location between the low- and \ha\ disk suggests that the \ha\ disk has had different mean migration directions and/or initial age-metallicity gradient in the star-forming gas, compared to the \lowalpha disk.
Unlike the \lowalpha disk where the turning points are at lower metallicity moving away from the Galactic center, suggesting an initial negative metallicity gradient in the gas, the \ha\ disk has age-[Fe/H] turning points roughly at the same metallicity at all locations across the Galaxy. This suggests the absence of any  initial metallicity gradient in this population.

The age-[Fe/H] turning point in the \ha\ disk does suggest that radial migration is a relevant evolutionary process in the \ha\ disk too. The empirical role and details of radial migration have been characterised to date only for the \lowalpha disk \citep[e.g.][]{Frankel2018, Frankel2019}. However, the role of radial migration in the \ha\ disk has been recognised really only within simulations \citep[eg][]{Roskar2008, Minchev2012, Buck2020,Khoperskov2020}.

One peculiar feature of Figure 8 is the reversal of the turning point in the \ha\ disk for 7 kpc $<$ R $<$ 9 kpc and $|z|<$ 0.5 kpc as well as the reversal of that of the \lowalpha\ disk at large $|z|$.
One explanation is that the method of using a line to distinguish the high- and \lowalpha disk is inappropriate. 
\cite{Ratcliffe2020} suggested a clustering algorithm to separate the high- and \lowalpha\ disk which resulted in a very different separation compared to that of a line. 
We will discuss evidence that this formalised method of assigning stars to groups using their chemical similarity may be preferable than a by-eye division in the Appendix.

\subsection{The intrinsic dispersion of elements around their age-abundance trends}\label{subsection:trends}
\begin{figure*}[htp]
    \centering
    \includegraphics[width=0.7\textwidth]{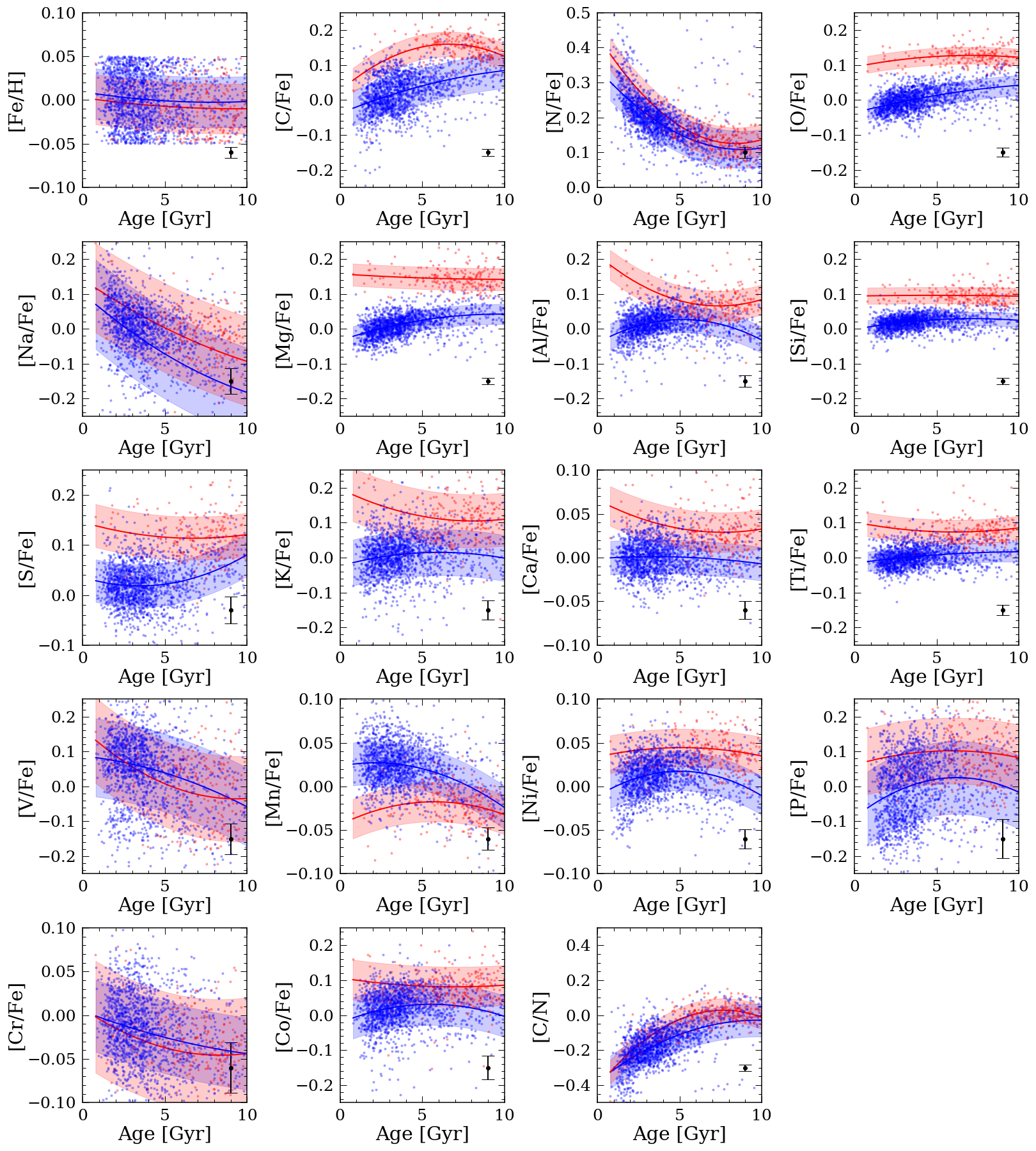}
    \caption{The age-abundance distributions and best-fit second-order polynomial relations for 16 elements as well as metallicity and [C/N] for the 1,651 \lowalpha and 224 \ha\ disk red clump stars, with ages determined from \texttt{The Cannon}.
    The shaded area shows the 1-$\sigma$ dispersion of the trends, and the typical error bars are shown in the bottom right corner.
    Note that the $y$-axis scales are different to show the details in the age-abundance relations.
    It is clear that the high- and \lowalpha disks have different age-abundance trends for most elements besides [C/N], [N/Fe], [V/Fe], and [Cr/Fe] ([Na/Fe] is excluded due to high systematic uncertainty).
    The age-metallicity relations are similar between the two disks ensure the differences in their relations are not caused by the metallicity.
    }
    
    \label{fig:abund_rc}
\end{figure*}

In previous sections, we looked at the overall age distribution for different spatial and chemical bins as well as the age-metallicity relation separated by the two disks.
We now examine the enrichment channels for the \ha\ and \lowalpha disk. Specifically we look to see if they are similar or show marked differences.

Since the metallicity of a star can significantly impact the element abundances \citep[see][]{Jofre2019}, we constrain our analysis to a narrow range in [Fe/H].
We look into detailed age-element abundance relations at a single reference solar metallicity, [Fe/H] = 0. We examine these relations for 16 elements (C, N, O, Mg, Al, Si, S,K, Ca, Ti, V, Mn, Ni, P, Cr, Co) observed by APOGEE. 
They mostly belong to the following nucleosynthetic families --- Iron-peak: V, Mn, Cr, Ni, Co, Sc; $\alpha$-element: O, Mg, Si, Ti,  Ca; Odd-z: Al, K, P and light: C, N 

We measure the age-abundance trends for each element at solar metallicity and calculate the intrinsic dispersions  (\sigmaint) around these relations \citep[][]{Ness2019}. That is, the scatter around the age-individual abundance trends not accounted for by the measurement errors.

Previous work has done this for the \lowalpha disk \citep[eg][]{Bedell2018,Ness2019} and a small \sigmaint\ has been found for almost all the elements. Recent work using GALAH has found small \sigmaint\ combining both high- and \lowalpha\ populations together \citep{Hayden2020, Sharma2020} In this section, we will investigate and compare the trends in the age-individual element relations and \sigmaint\ for the low- and \ha\ disk.

We use two sets of stars for this (i) the benchmark set of asteroseismic stars identified as red clump stars with 1,261 \ha\ and 3,173 \lowalpha stars, with typical age errors of 0.88 Gyr and (ii) the red clump stars identified as per section \ref{sec:datamethod}, with 5,290 \ha\ and 16,784 low alpha stars respectively, with typical age errors of 3.4 Gyr. 

To do so, we selected stars with asteroseismic ages from \cite{Pinsonneault2018} as well as red clump stars with solar metallicity ($|$[Fe/H]$| < $ 0.05 dex).
We also restricted to stars with metallicity error $<$ 0.03 dex.
We excluded 11 \ha\ and 12 \lowalpha stars from the asteroseismic sample and 46 \ha\ and 114 \lowalpha red clump stars with age $>$ 10 Gyr. 
The relations for stars $>$ 10 Gyr are fairly flat (also in part presumably due to large age errors) and here we characterise the gradient where a trend is present (see Figure~\ref{fig:abund_rc} and \citealt{Ness2019}).
We also excluded stars that have $\chi^2 > 10,000$ (with $\bar{\chi}^2 > 1.17$) between the spectra model by \texttt{The Cannon} and the real spectra.
This leaves us with 642 \lowalpha disk stars, 53 \ha\ disk stars with asteroseismic ages and 1,650 \lowalpha disk stars, 224 \ha\ disks stars with ages determined from \texttt{The Cannon}.

To make sure there are no systematic temperature dependencies for the abundances, we examined the abundances of these stars versus stellar age, as a function of temperature. We see no temperature gradients along the abundance axis, with the exception of the element V.
This could indicate bias in the measurement of V, as a result, we excluded this element in analyzing the intrinsic dispersion and the slope of the age-abundance relations. 
For each element we determine the age-abundance relations, using both samples of stars. 

The results for the solar metallicity red clump stars are shown in Figure~\ref{fig:abund_rc}.
We show lines fit using a second order polynomial, to quantify these age-abundance relations. 
The blue and red lines show the fit to the low- and \ha\ stars, respectively. The shaded region represent the total dispersion around the relations, and the typical error bar for each element is shown in the bottom right corner.
In general, the average abundances for the \ha\ disk are higher than those of the \lowalpha\ disk and the relations between the two disks are different apart from [C/N], [N/Fe], [V/Fe] and [Cr/Fe].
The differences in the age-[Mg/Fe] trends for the high- and \lowalpha disk is also observed in \cite{Kochukhov2021}. 
The [Fe/H]-age relations are similar between the two disks ensure the differences in the other abundance relations are not caused by the difference in metallicity.

We also calculated the slopes of these trends by fitting straight lines through the age-abundance relations for the red clump stars. 
However, the relations in Figure~\ref{fig:abund_rc} suggests the linear relation is in log(age)-abundance space ([X/Fe]=a log(age)+b), but we calculated the slope in linear age space to compare our results with literature values from \cite{Bedell2018, Ness2019}.

\begin{figure*}[htp]
    \centering
    \includegraphics[width=\textwidth]{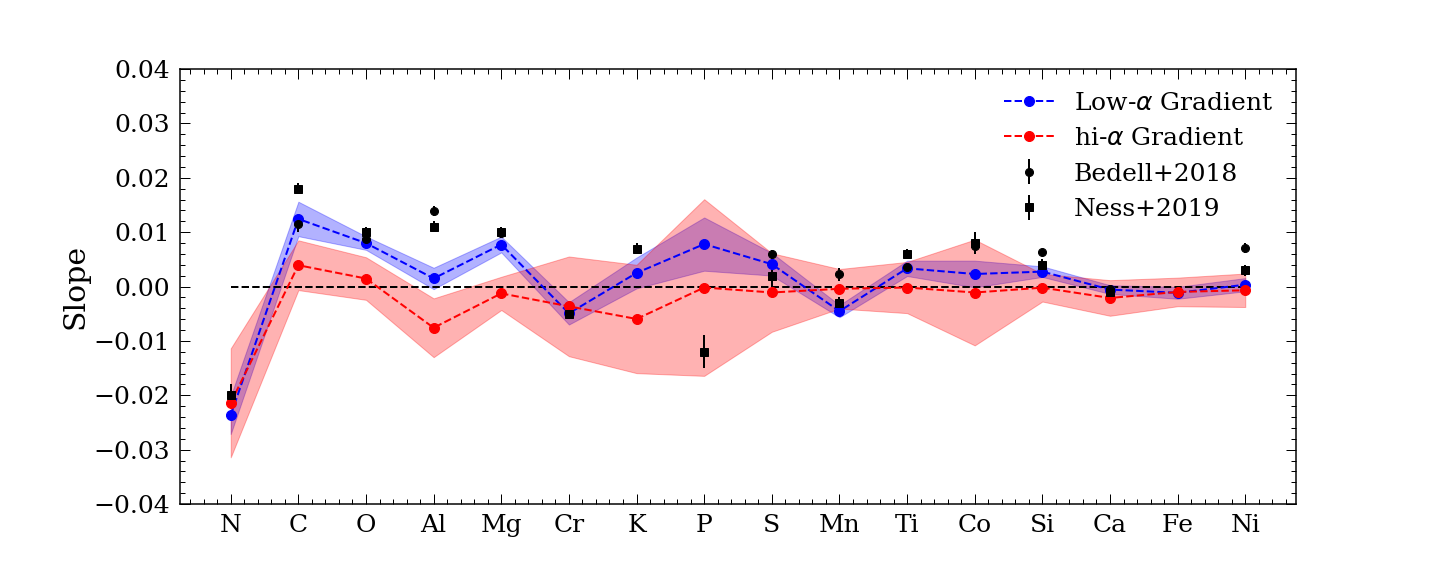}
    \caption{Slopes of the detailed age-element abundance relations for solar metallicity stars [Fe/H] = 0 $\pm$ 0.05 dex, ordered by the average absolute slope of the high- and \lowalpha disk (from largest to smallest). The shaded area represents the $1-\sigma$ confidence of the measurement of the slope, determined by calculating the $1-\sigma$ dispersion in the best-fit slope from 500 re-draws of the measurements from their age and individual abundance errors, for each element.
    The slope is calculated using a straight line. Results from the main sequence study of \citet{Bedell2018} and \citet{Ness2019} are included for comparison. Note these two studies are for the low-$\alpha$ stars only.
    }
    \label{fig:slopes}
\end{figure*}

The results are shown in Figure~\ref{fig:slopes}, in which the red dots represent the slopes for the trends in the \ha\ disk, the blue dots represent those for the \lowalpha disk, the black circles show the results from \cite{Bedell2018}, and the black squares show the results from \cite{Ness2019}.
The elements are arranged so that the average absolute slope of the two sequence decreases towards the right of the $x$-axis.
We excluded Na and V from our calculation as we suspect systematic bias in the measurements as previously discussed.

The uncertainties (shown as the shaded area) were measured by perturbing each point within its uncertainties both in age and abundance, fitting a new line each time.
We recalculated the slope 500 times weighting in the uncertainties of the data, and the uncertainty was then determined by the standard deviation of all these 500 measurements. 

We also tested how the slope varies across Galactic radius by calculating the age-abundance relation slope in bins of 1 kpc width, between R = 8 kpc to 13 kpc. In doing this, we found no significant spatial variation of the age-abundance slopes. Thus, we conclude the age-individual abundance relations are global across the disk, conditioned on [Fe/H]; note that they change across [Fe/H] so would presumably change spatially if not examined at this fixed metallicity \citep[see][]{Ness2019}.

The slopes between the high- and \lowalpha disk stars are similar for some elements (e.g. N, V, Cr, Mn) and quite different for others (e.g. Al, Mg, O, C).
We expected the slopes for the \lowalpha disk to be slightly dissimilar from the results shown in \cite{Bedell2018} since they used dwarf stars around solar temperature as opposed to giant stars in this work. Differences may hint at the contribution of stellar versus galactic chemical evolution. Similarly, the slopes are very similar to those reported in \citet{Ness2019}, comparing the low-$\alpha$ disk results. 

We then calculated the \sigmaint\ using the method described in \cite{Ness2019},
\begin{equation}
    \sigma_{int}^2 = \sigma_{tot}^2 - \sigma_{measure}^2,
\end{equation}
where $\sigma_{tot}$ is the total dispersion, measured by calculating the dispersion around the best-fit 2$^{nd}$ order polynomial, and $\sigma_{measure}$ is the measurement dispersion.
The measurement dispersion is estimated by perturbing the abundance and age of each point within their uncertainties and then measure the dispersion around the original determined 2$^{nd}$ order polynomial.
We perform this 100 times and $\sigma_{measure}$ was then taken to be the standard deviation of the 100 dispersion measurements.

\begin{figure*}[htp]
    \centering
    \includegraphics[width=\textwidth]{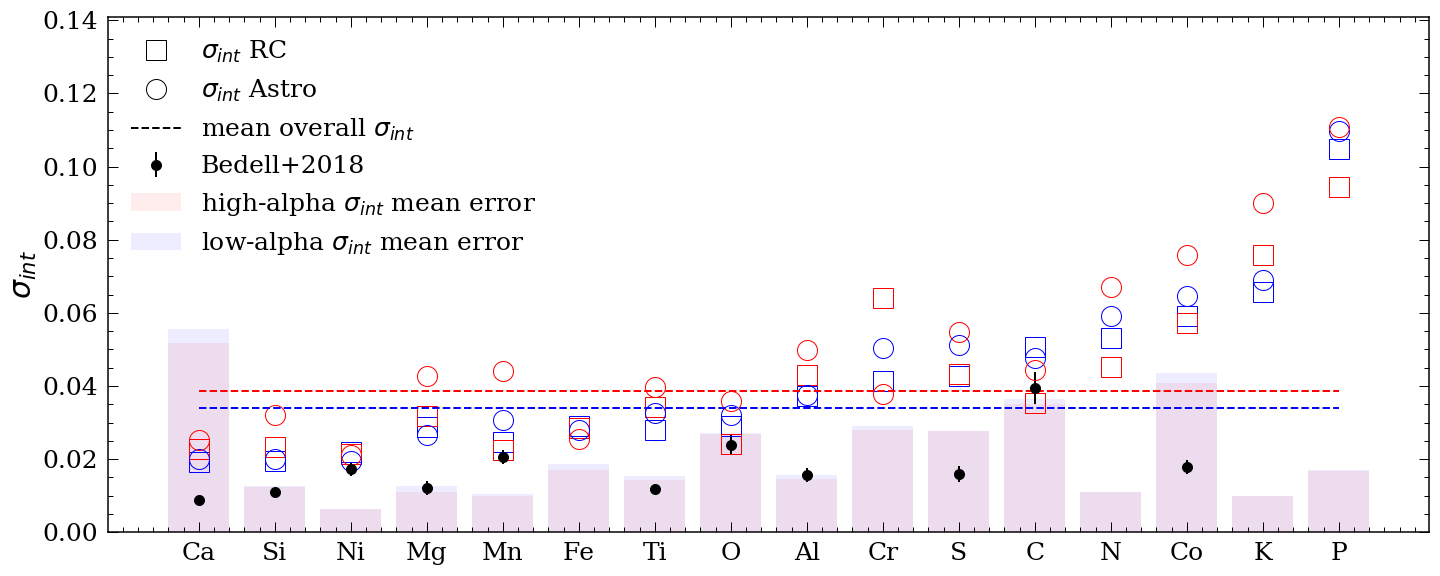}
    \caption{Intrinsic dispersion at solar metallicity, [Fe/H] = 0 $\pm$ 0.05 dex, for each element in the high- (red) and \lowalpha (blue) disk for both stars with asteroseismic ages from \cite{Pinsonneault2018} (circles) and ages determined from \texttt{The Cannon} (squares). 
    The black dots are results from \cite{Bedell2018}; the bars show the mean abundance error for the high- (red) and \lowalpha (blue) red clump stars; and the dashed lines show the median intrinsic dispersion of the elements, which is 0.039 and 0.035 for the high- and \lowalpha sequence respectively.
    These indicate that there is no obvious correlation between measurement error on the elements and the calculated intrinsic dispersion. 
    Elements are ordered by the average dispersion of the asterosismic \lowalpha\ stars and the red clump \lowalpha stars. }
    \label{fig:disp}
\end{figure*}

Figure~\ref{fig:disp} shows the intrinsic dispersion for the stars with asteroseismic ages from \cite{Pinsonneault2018} (circles) and ages determined from \texttt{The Cannon} (squares), separated by the high- (red) and \lowalpha (blue) disk.
The dashed lines show the median dispersion for the high- (red) and \lowalpha (blue) disk and the bars are the mean abundance error for the high- (red) and \lowalpha (blue) red clump stars.
Our intrinsic dispersion measurements are very similar to those of \cite{Bedell2018} (and \citet{Ness2019} who also found consistent results), with the exception of V, Na, and Co. 
The slopes of the age-abundance relations for the high- and \lowalpha disk are very similar, and are not correlated with the intrinsic dispersion.

We found very similar intrinsic dispersion between the high- and the \lowalpha disk, with a median of 0.039 dex and 0.035 dex, respectively. 
Note \cite{Vincenzo2021} also calculated the intrinsic dispersion for [Mg/Fe] and found a value of $\sim$ 0.04 dex for both the high- and \lowalpha sequence. 
The slightly higher dispersion in the \ha\ disk might simply indicate a faster rate of enrichment which leads to more variability in the range of each element's abundance at any given age, at fixed metallicity [Fe/H].

The low intrinsic dispersion that we report here suggests we should be able to determine ages using the detailed age-element abundance relations for both the low- and the \ha\ disk \citep[see also][]{Hayden2020, Sharma2020}.
Further, our result hints at universal chemical enrichment processes that have given rise to the abundance distributions of the high- and \lowalpha disk.
The small variation in the intrinsic dispersions indicate subtle differences in the nucleosynthesis processes that are taking place.

\begin{figure*}[htp]
    \centering
    \includegraphics[width=\textwidth]{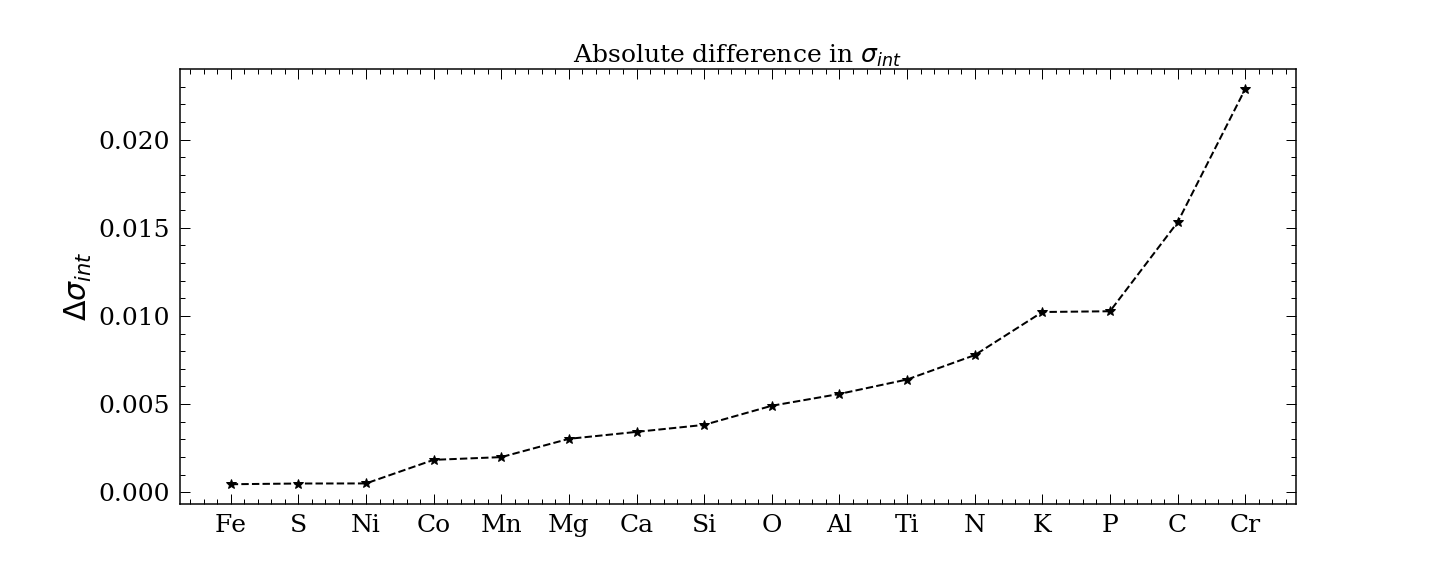}
    \caption{Absolute difference of the intrinsic dispersion around the age-abundance relations \sigmaint\ between the high- and \lowalpha disk for the red clump stars at solar [Fe/H] = 0 $\pm$ 0.05 dex. These are ordered by increasing difference in the intrinsic dispersion. }
    \label{fig:disp_diff}
\end{figure*}

Figure~\ref{fig:disp_diff} shows the absolute difference of \sigmaint\ between the high- and \lowalpha disk for the red clump stars.
Elements with small $\Delta$\sigmaint\ are mostly iron-peak elements, and those with large $\Delta$\sigmaint\ are mostly odd-z elements.

\section{Discussion \& future work}\label{sec:dis}

Large spectroscopic surveys such as Apache Point Observatory Galactic Evolution Experiment (APOGEE) \citep{Majewski2017}, Large Sky Area Multi-Object Fibre Spectroscopic Telescope (LAMOST) \citep{LAMOST}, GALactic Archaeology with HERMES (GALAH) \citep{Silva2015, Buder2019} and time-domain surveys such as \kepler\ \citep{Borucki2010} and \tess\ \citep{TESS} are observing hundreds of thousands of stars.
These surveys enable us to test galaxy formation and evolution mechanisms as well as channels of element production. 

By using the APOGEE DR16 spectra, measurements of frequency spacing between $p$-modes, \dnu, and period spacing of the mixed $g$ and $p$ modes, \dP, from \cite{Vrard2016}, as well as estimation of ages from the second APOKASC catalog \citep{Pinsonneault2018} and \teff, \logg, metallicity [Fe/H], and [Mg/Fe] from APOGEE's DR16, we constructed an age catalogue for 64,317 stars in APOGEE derived using The Cannon with $\sim$ 1.9 Gyr uncertainty across all ages (APO-CAN stars) as well as a red clump catalogue of 22,031 stars with a contamination rate of 2.7\%.

Combining our catalogs and 16 element abundances (C, N, O, Mg, Al, Si, S, K, Ca, Ti, V, Mn, Ni, P, Cr, Co) from \cite{DR162020}, we concluded several similarities and differences between the high- and \lowalpha disk:
\begin{enumerate}
   \item Similarities:
   \begin{itemize}
     \item Both disks show evidence for radial migration and upside-down formation (Figures~\ref{fig:agedist}~\&~\ref{fig:Feage}).
     \item Intrinsic dispersions around the age-abundance relations are small, highlighting the universality of temporal chemical enrichment pathways (Figure~\ref{fig:disp}).
     \item Given large numbers of stars as in our study, the high and \lowalpha\ disk can be described within a single framework as an ensemble of temporal and spatial populations across chemical space that underlie a global disk distribution. Presumably the gradients in the mean age, dynamics and spatial extent (and higher order moments of these distributions) afford very strong constraints on the Galaxy's formation and evolution (Figure~\ref{fig:age_alpha}, \ref{fig:disp}).
   \end{itemize}
   \item Differences:
   \begin{itemize}
     \item Our analysis suggests that the \ha\ disk had no initial metallicity gradient, where as the \lowalpha disk formed with a gradient in place (Figure~\ref{fig:agedist}).
     \item The \ha\ disk is old and has a larger vertical reach and narrower radial expanse than the young \lowalpha disk, which has a shorter vertical reach and extends over a larger radial range (Figure~\ref{fig:age_space}, \ref{fig:age_alpha}).
     \item There are differences in the age-[Fe/H] relations for the high- and \lowalpha sequences across (R,z). These suggest distinct rates and/or directions of radial migration and/or different initial metallicity gradients with Galactic radius (Figure~\ref{fig:Feage}).
     \item Most of the age-individual abundance relations for the high- and the \lowalpha\ disk (Figure~\ref{fig:abund_rc}) show differences, with some exceptions. Some elements, such as N, show near-identical relations. Other elements, like Mg and Al, have quite different slopes (see Figure 10). 
     \item Although some elements have identical intrinsic dispersions around the age-individual abundance relations for the high and low-$\alpha$ disk, on average this is abour 10\% larger in the \ha\ disk. This hints at either different star formation efficiencies/rates or the level of mixing of chemical elements in  the star-forming gas  (Figure~\ref{fig:disp}, \ref{fig:disp_diff}).

   \end{itemize}
\end{enumerate}

\subsection{Comparing with simulations/analytic models}

In this paper, we provide a population study of the similarities and differences between the high- and \lowalpha disks using APOGEE DR16 and \gaia\ data.
In this section, we will briefly summarise our results in the context of formation mechanisms seen in simulations and in analytic models.

Figure~\ref{fig:agedist} examines the global age distribution and the metallicity skewness of the two disks. The skewness gradient for all stars is seen in the simulations of \cite{Agertz2020}, suggesting radial migration.

Figure~\ref{fig:age_space} revealed the age distribution of stars in small spatial bins. The lack of age gradient in the \ha\ disk is seen in the simulations of \cite{Agertz2020}, and the strong age gradient in the \lowalpha disk resembles that of the simulations in  \cite{Buck2020}. Furthermore, the flat age gradient in the \ha\ disk points towards a fast formation mechanism which is in line with the recent findings of \citet{DiMatteo2020}.

Figure~\ref{fig:age_alpha} showcased the age distribution of stars in different chemical cells together with their spatial distribution across the disk. In general, older stars have shorter scale length and extend to larger heights above the plane compared to younger stars. This is in line with early star formation happening in a compact turbulent gas disk \citep[e.g.][]{Buck2020b} while the low-$\alpha$ disk forms later \citep[e.g.][]{Lian2020}. 
We found the largest age dispersion appears at the most metal rich and $\alpha$ poor region (see Figure~\ref{fig:age_chem_disp}), supporting the analytic two in-fall model described in \cite{Spitoni2021}.
Furthermore, the different age-abundance relations for low- and high-$\alpha$ stars shown in Fig. \ref{fig:abund_rc} might hint towards different formation mechanisms/time scales or star formation efficiencies and/or the level of mixing of chemical elements in the star-forming gas. Similar conclusions have also been made by \citet{Nissen2020} using HARPS data.

Figure~\ref{fig:Feage} shows the age-metallicity relations at different locations in the disk in (R, z). These results suggest that radial migration has been significant in both disks. This result is supported by \cite{Buck2020,Khoperskov2020} but are  in tension with recent formation scenarios without strong radial migration \citep{Khoperskov2021}.

\subsection{Limitations}
We did not take into account the selection function, this means we are not able to calculate the scale height or scale length of the two disks. 
However, we argue that the major results will not change because of the simplicity of the APOGEE selection function.

We only examined stars with solar metallicity, [Fe/H] = 0 for the detailed age-abundance trends. However, by testing the intrinsic dispersion for stars with metallicity centered around -0.3 dex, we concluded the intrinsic dispersion does not vary much, although the age-abundance relations are different for stars with different overall metallicity, [Fe/H].

\section{Conclusions}
With such a large and detailed benchmark comparison of the age-abundance relations across the chemically defined high- and \lowalpha\ disk, these results can potentially constrain nucleosynthesis channels across broadly different chemical regimes. 
While the similarity across chemical space in the intrinsic dispersions in the age-abundance relation indicates the universality of chemical enrichment, the small element-dependent differences we see are relevant in informing metallicity dependent yields, and the impact of different star formation rates and environment (Buck et al. in prep.). 

Using the distributions of ages of stars across chemical and spatial cells, we hope to provide the empirical data to distinguish between different formation scenarios for the Galaxy, by e.g., comparing to cosmological simulations. Moving forward, we expect a powerful analysis tool to constrain the origin of the bi-modality in the disk will be to examine the age, dynamical and spatial properties of stars as a function of their chemical distributions. To first order, simulations that reproduce a bi-modality in the \alphafe\ plane must also reproduce the results we see in Figure 7, given equivalent sampling, to reflect the formation channel(s) that underlie the Milky Way's current set of observed properties. To second order, Figure 8 is demonstrative of the strength of evolutionary processes like radial migration across different chemical (and correspondingly, initial spatial) spaces. Next generation surveys \citep[e.g.][]{Kollmeier2017} will enable this analysis to be taken to the next level by completing this exploration over a vast expanse of the disk into the bulge and to sample the disk finely across its temporal and spatial variables, within true mono-abundance chemical populations.

\appendix \label{sec:append}
\renewcommand{\thesubsection}{\Alph{subsection}}
\counterwithin{figure}{subsection}
\counterwithin{table}{subsection}

\subsection{Predicting stellar ages with \texttt{Astraea}}
We also tried to predict stellar ages estimated using \texttt{The Cannon} with \texttt{Astraea} \citep{Lu2020}\footnote{Avaliable at \url{https://astraea.readthedocs.io/en/latest/}.}.
\texttt{Astraea} uses Random Forest, a machine learning algorithm, to predict label from features.

We performed a simple cross-match between APOGEE and the \gaia\-\kepler\ cross-match catalog\footnote{Avaliable at https://gaia-kepler.fun.} and found 3,417 stars with the measurements we needed.
The features we trained on are all the 17 element abundances from APOGEE, metallicity, and all the \gaia\ parameters.
The important features are listed in table~\ref{tab:astraea}.

\begin{table}
    \centering

    \begin{tabularx}{0.5\textwidth}{l|X|X}
    
    \hline
    Feature name & Description & source   \\
    \hline
    \hline
    N\_FE & [N/Fe] & APOGEE \\
    \hline
    C\_FE & [C/Fe] & APOGEE \\
    \hline
    MG\_FE & [Mg/Fe] & APOGEE \\
    \hline
    TI\_FE & [Ti/Fe] & APOGEE \\
    \hline
    AL\_FE & [Al/Fe] & APOGEE \\
    \hline
    ALPHA\_M & [$\alpha$/Fe] & APOGEE \\
    \hline
    TEFF\_SPEC & spectroscopic temperature & APOGEE \\
    \hline
    RV\_CCFWHM & radial velocity & APOGEE\\
    \hline
    radius\_val & radius value & \gaia\ \\
    \hline
    pmdec & proper motion in dec & \gaia\ \\
    \hline
    pmra & proper motion in ra & \gaia\ \\
    \hline
    parallax & parallax & \gaia\ \\
    \hline
    b & Galactic latitude & \gaia\ \\
    \hline
    l & Galactic longtitude & \gaia\ \\
    \hline
    Log(g) & \logg & \texttt{The Cannon}\\
    \hline
    r\_est & estimated distances & \cite{Bailer2018}\\
    \hline
    \end{tabularx}
    \caption{Stellar parameters used to train \texttt{Astraea} to get stellar ages.}
    \label{tab:astraea}
\end{table}

We trained on 80\% of the data and predicted the stellar ages for the rest of the 20\%.
We were able to predict these ages from \texttt{The Cannon} with a median relative error of 15\%.
Figure~\ref{fig:predictage} shows the ``{\it gini}'' importance (ranges from 0 to 1, where 1 being the most important) of the 18 most important feature in predicting these ages. 
This importance can be determined by calculating the mean decrease in impurity (MDI), which indicates whether a single feature alone can predict the outcome.
For example, if one can predict the stellar ages of a star just by the effective temperature, then the {\it gini} importance for the effective temperature will be 1.

\setcounter{figure}{0}
\begin{figure*}[htp]
    \centering
    \includegraphics[width=\textwidth]{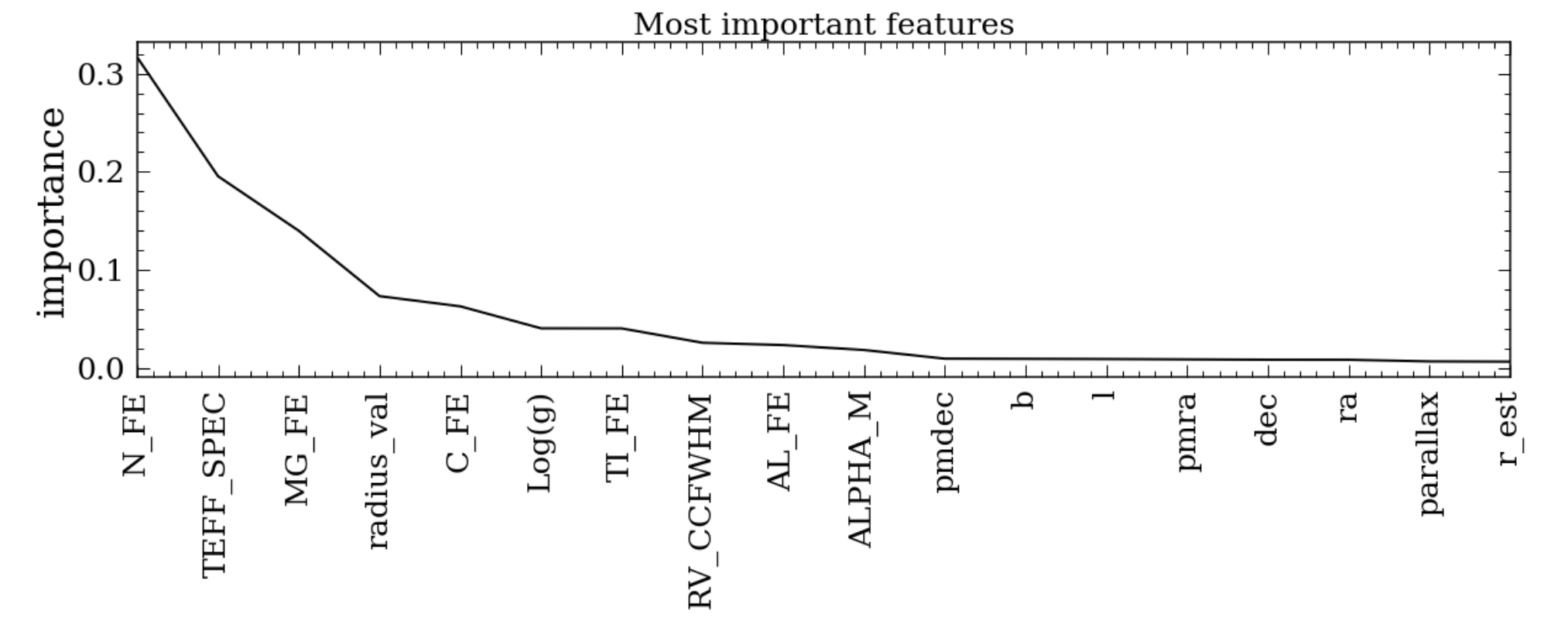}
    \caption{The ``{\it gini}'' importance of the features used to predict stellar ages using \texttt{Astraea}.
    This indicates that we are able to predict stellar ages with only a few stellar parameters.}
    \label{fig:predictage}
\end{figure*}

It is not surprising [N/Fe] is the most important feature in determining the stellar ages, as age-[N/Fe] relation has one of the steepest slopes (see Figure~\ref{fig:slopes}) and smallest intrinsic dispersion (see Figure~\ref{fig:disp}).
The fact that [Na/Fe] is not important in determining the ages despite the fact that it has the steepest slope further indicates that there is anomaly in this abundance measurement.
The importance of \alphafe\ is relatively low, suggesting the two disks experience similar enrichment process.
The distribution of importance is relatively spread-out (in that several features are relatively important to predict ages) suggesting determining stellar ages is complicated.
It also suggests stellar ages are indeed related to a sum of complicated stellar processes such as nucleosynthesis, gravity, kinematics.

Even with a spread in importance, it is still striking that we were able to predict stellar ages within 20\% uncertainty with only a handful of stellar parameters.
This means we should be able to estimate stellar ages for a large number of field stars fairly straightforwardly with large spectroscopic survey such as LAMOST and GALAH. 

\subsection{Separating the high- and \lowalpha disk with a clustering algorithm}
\begin{figure}[htp]
    \centering
    \includegraphics[width=\textwidth]{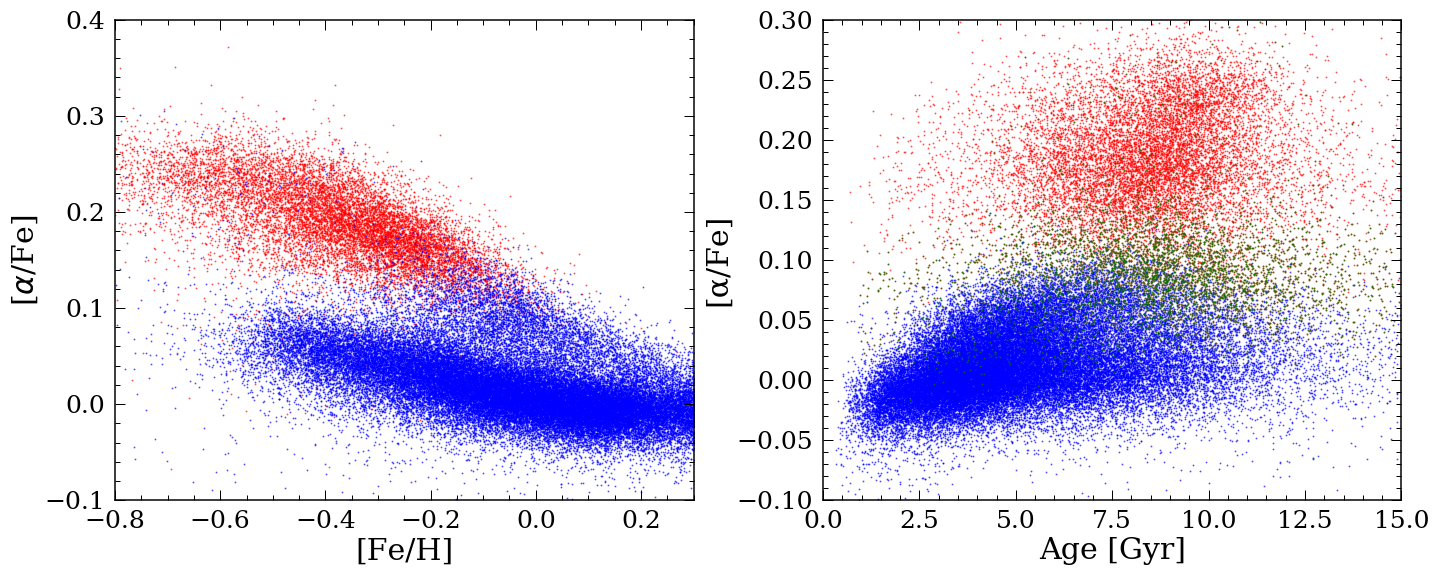}
    \caption{The high- (red) and \lowalpha (blue) disk separated using ward hierarchical clustering.
    We were able to reproduce the result in \cite{Ratcliffe2020}.
    The green points are stars that were considered \ha\ disk stars with our separation but classified as \lowalpha disk stars with the clustering algorithm. 
    The separation of the two disk in the age-\alphafe\ plain (right) favors the clustering algorithm.}
    \label{fig:clustering}
\end{figure}

In this paper, we separated the high- and \lowalpha disk with an ad-hoc straight line.
However, it is not clear whether this is the most appropriate way to separate the two disks. Particularly since the [$\alpha$/Fe] and metallicity, [Fe/H] only represent two of the chemical dimensions of a larger chemical space.
\cite{Ratcliffe2020} suggested a clustering algorithm approach to deconstruct stars of the disk, using the 19 dimensions of available APOGEE abundances. In their work, they reported that a small group of the \ha\ stars was more stronly associated to the  \lowalpha disk than the high.
We performed the same ward hierarchical clustering as per \citet{Ratcliffe2020}, using \texttt{sklearn} \citep{scikit-learn} with the same 17 elements and [Fe/H] and found the same two-cluster projection in the [$\alpha$/Fe]-[Fe/H] as in that work. 
The clustering result is shown in Figure~\ref{fig:clustering}.
The green points are stars that are considered \ha\ stars with the line separation used in this paper, but are classified as \lowalpha stars in \cite{Ratcliffe2020}.

One line of independent evidence that these nominallly \ha\ disk stars could be associated with the \lowalpha disk more than the high, is from the age-metallicity relations.
As pointed out in the section \ref{subsection:agemetal}, we were not able to explain some of the features in the age-metallicity relations for the \ha\ disks.
For example, the reversal of the turning point in age for the spatial bin of 7 $<$ R $<$ 9 and $|z| <$ 0.5. 
Using the high and low-$\alpha$ disks defined by hierarchical clustering, we are able to extract what look to be clearer and more distinctly different relations for the \ha\ disk stars across the galaxy.

\begin{figure*}[htp]
    \centering
    \includegraphics[width=\textwidth]{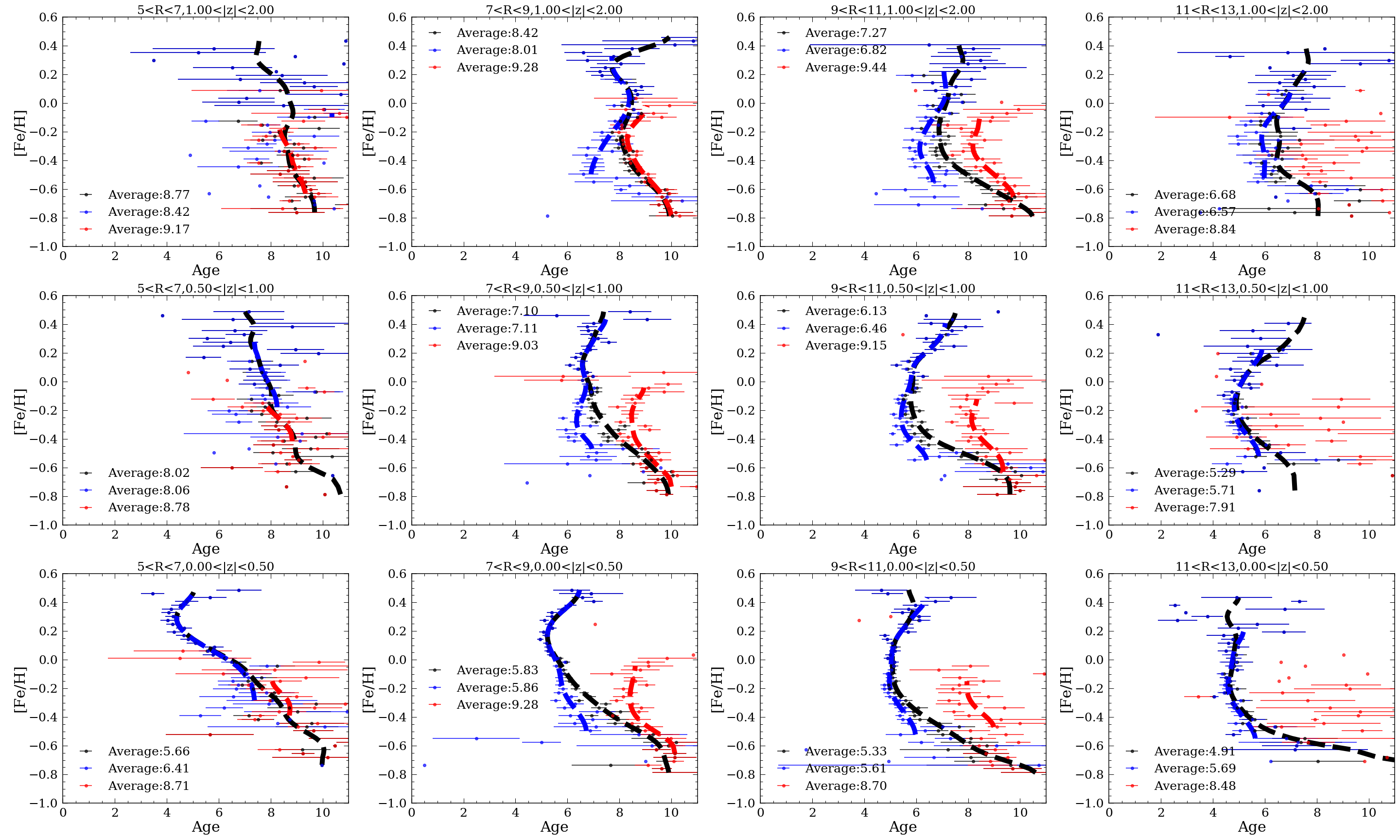}
    \caption{Same as figure~\ref{fig:Feage} but the high- and \lowalpha disks are determined by the clustering algorithm.}
    \label{fig:Feage_cluster}
\end{figure*}

We also tried using more than 2 clusters and we see that there only exists two distinct modes, meaning the age-metallicity relation of the third (and above) cluster overlaps with that of the \lowalpha disk.

However, note that clustering algorithms, similarly to by-eye designations, do not offer a ``best'' solution. They are subject to algorithmic choices. These results should be interpreted with caution and used to inform what varies under different analysis choices.

\subsection{Additional graphs}
\begin{figure*}[htp]
    \centering
    \includegraphics[width=\textwidth]{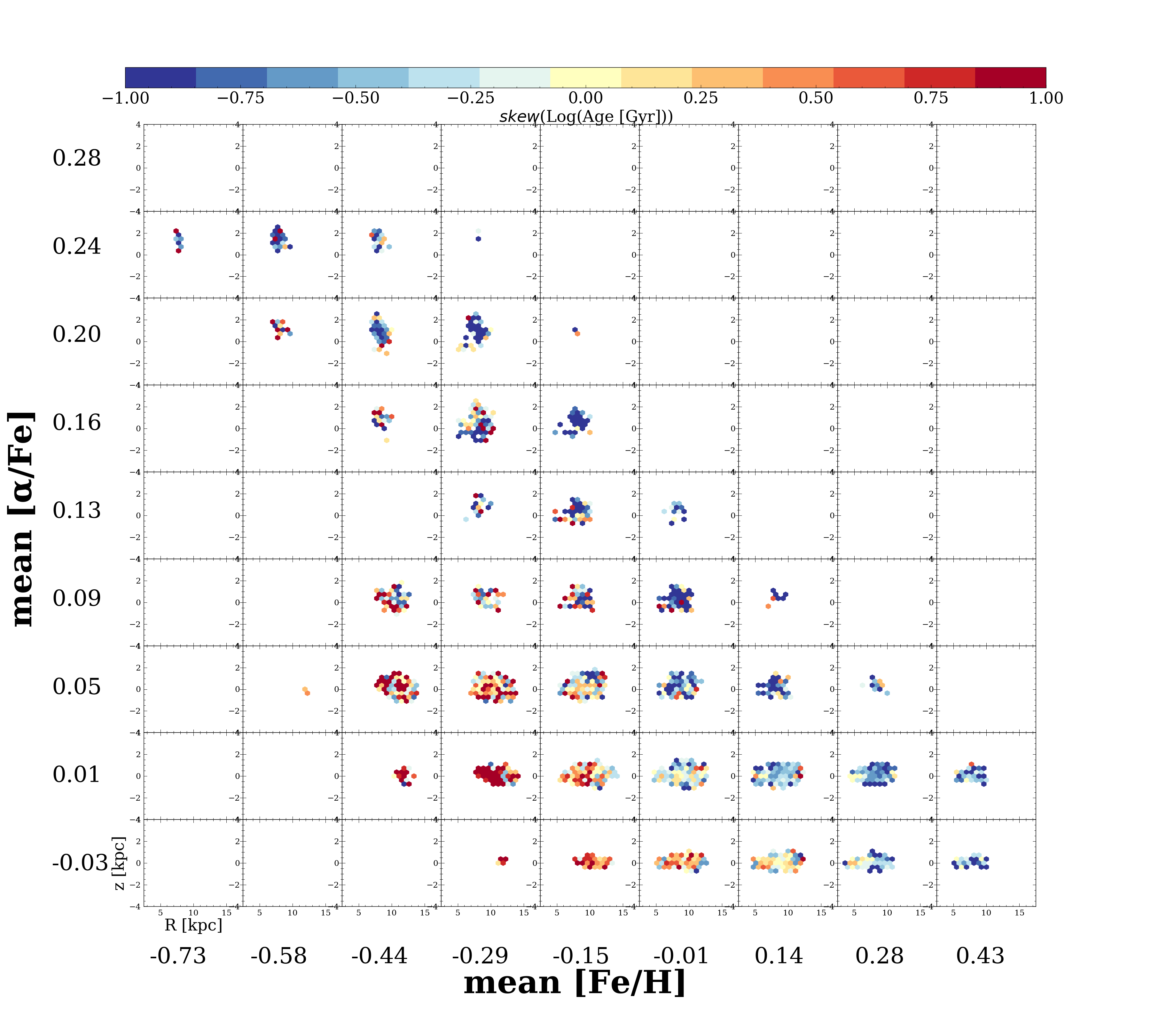}
    \caption{Same as Figure~\ref{fig:age_alpha} but plotting the skew in age. Only included bins with $>$ 10 stars.}
    \label{fig:age_chem_skew}
\end{figure*}

\begin{figure*}[htp]
    \centering
    \includegraphics[width=\textwidth]{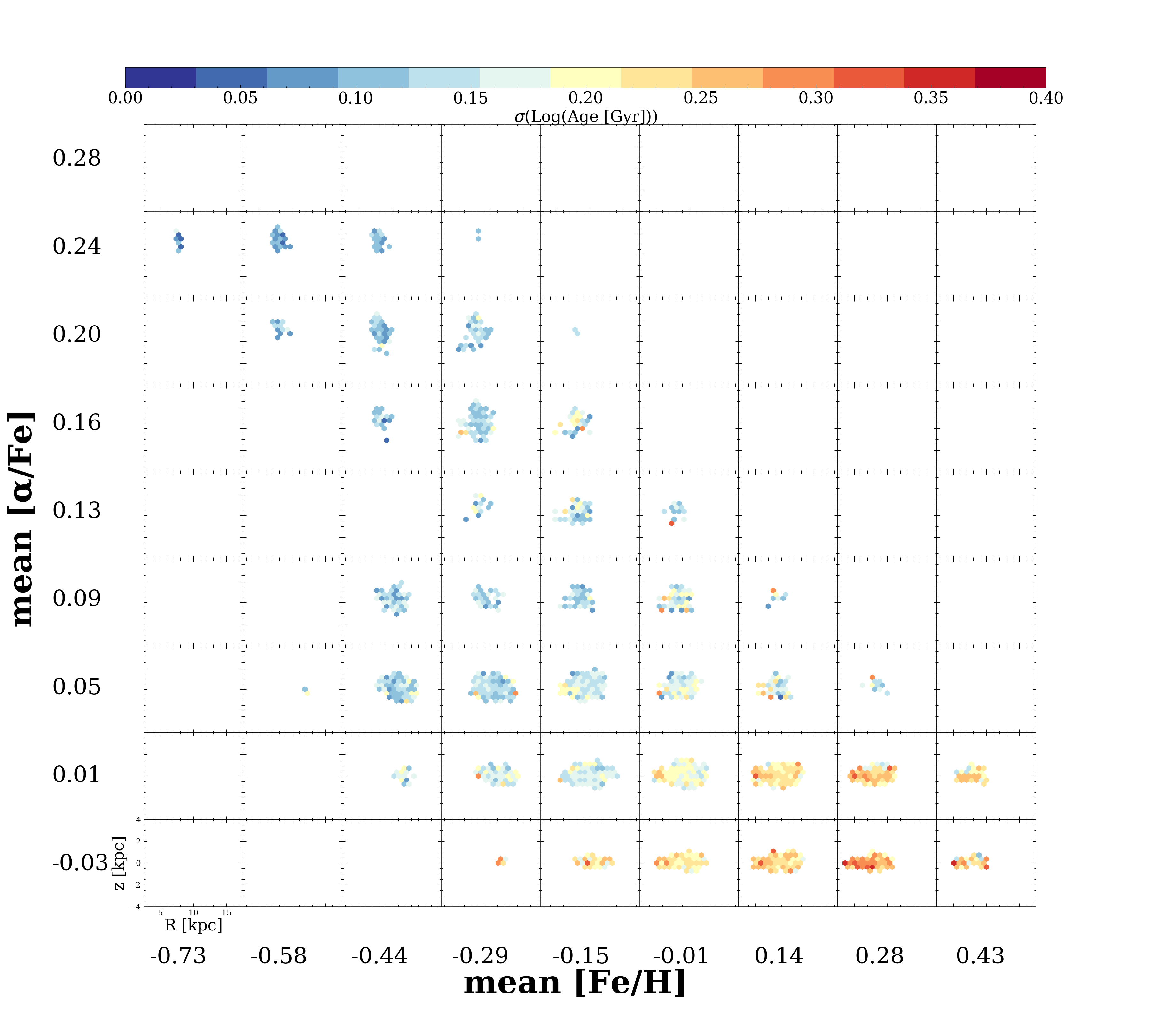}
    \caption{Same as Figure~\ref{fig:age_alpha} but plotting the standard deviation in age. Only included bins with $>$ 10 stars.}
    \label{fig:age_chem_disp}
\end{figure*}

Figure~\ref{fig:age_chem_skew} and \ref{fig:age_chem_disp} show the skew and standard deviation of the ages in each bin using the \texttt{matplotlib.axes.Axes.hexbin} function.

\acknowledgments
We thank Adrian Price-Whelan and Kate Daniel for helpful discussions.

JCZ is supported by an NSF Astronomy and Astrophysics Postdoctoral Fellowship under award AST-2001869.

Melissa Ness is supported in part by a Sloan Fellowship. 

This work made use of the gaia-kepler.fun crossmatch database created by Megan Bedell.

This paper includes data collected by the {\it Kepler} mission. Funding for the {\it Kepler} mission is provided by the NASA Science Mission directorate.

This work has made use of data from the European Space Agency (ESA) mission
{\it Gaia} (\url{https://www.cosmos.esa.int/gaia}), processed by the {\it Gaia}
Data Processing and Analysis Consortium (DPAC,
\url{https://www.cosmos.esa.int/web/gaia/dpac/consortium}). Funding for the DPAC
has been provided by national institutions, in particular the institutions
participating in the {\it Gaia} Multilateral Agreement.

Funding for the Sloan Digital Sky 
Survey IV has been provided by the 
Alfred P. Sloan Foundation, the U.S. 
Department of Energy Office of 
Science, and the Participating 
Institutions. 

SDSS-IV acknowledges support and 
resources from the Center for High 
Performance Computing  at the 
University of Utah. The SDSS 
website is www.sdss.org.

SDSS-IV is managed by the 
Astrophysical Research Consortium 
for the Participating Institutions 
of the SDSS Collaboration including 
the Brazilian Participation Group, 
the Carnegie Institution for Science, 
Carnegie Mellon University, Center for 
Astrophysics | Harvard \& 
Smithsonian, the Chilean Participation 
Group, the French Participation Group, 
Instituto de Astrof\'isica de 
Canarias, The Johns Hopkins 
University, Kavli Institute for the 
Physics and Mathematics of the 
Universe (IPMU) / University of 
Tokyo, the Korean Participation Group, 
Lawrence Berkeley National Laboratory, 
Leibniz Institut f\"ur Astrophysik 
Potsdam (AIP),  Max-Planck-Institut 
f\"ur Astronomie (MPIA Heidelberg), 
Max-Planck-Institut f\"ur 
Astrophysik (MPA Garching), 
Max-Planck-Institut f\"ur 
Extraterrestrische Physik (MPE), 
National Astronomical Observatories of 
China, New Mexico State University, 
New York University, University of 
Notre Dame, Observat\'ario 
Nacional / MCTI, The Ohio State 
University, Pennsylvania State 
University, Shanghai 
Astronomical Observatory, United 
Kingdom Participation Group, 
Universidad Nacional Aut\'onoma 
de M\'exico, University of Arizona, 
University of Colorado Boulder, 
University of Oxford, University of 
Portsmouth, University of Utah, 
University of Virginia, University 
of Washington, University of 
Wisconsin, Vanderbilt University, 
and Yale University.

Guoshoujing Telescope (the Large Sky Area Multi-Object Fiber Spectroscopic Telescope LAMOST) is a National Major Scientific Project built by the Chinese Academy of Sciences. Funding for the project has been provided by the National Development and Reform Commission. LAMOST is operated and managed by the National Astronomical Observatories, Chinese Academy of Sciences.

This research made use of Astropy,\footnote{http://www.astropy.org} a community-developed core Python package for Astronomy \citep{astropy:2013, astropy:2018}.

%

\vspace{5mm}
\facilities{Gaia, Kepler, APOGEE}


\software{Astropy \citep{astropy:2013, astropy:2018}, Numpy \citep{oliphant2006guide}, sklearn \citep{scikit-learn}, The Cannon \citep{Ness2015}, Astraea \citep{Lu2020}}




\bibliography{references.bib}{}
\bibliographystyle{aasjournal}



\end{document}